\documentclass[12pt]{article}

\usepackage[hyphens]{url}%

\usepackage[automake,xindy,acronym]{glossaries-extra}%

\usepackage{xfrac}%
\usepackage{bbold}%

\makeglossaries%

\GlsXtrEnablePreLocationTag{(page }{(pages }%

\newglossaryentry{AM}{%
  name=AM,%
  description={Arthur-Merlin. The set of decision problems that can be decided in polynomial time by an Arthur–Merlin protocol}%
}%
  
\newglossaryentry{BPP}{%
  name=BPP,%
  description={Bounded-error Probabilistic Polynomial-time . The class of decision problems that probabilistic Turing machines can solve in polynomial time with an error probability bounded by \sfrac{1}{3} for all instances of the problem}%
}%
  
\newglossaryentry{CC}{%
  name=CC,%
  description={Cloud Computing. The delivery over the internet of computing services, including data storage and computing power, on-demand with pay-as-you-go pricing (definition from NIST \cite{MellG2011}; other interpretations exist)}%
}%
  
\newglossaryentry{CRH}{%
  name=CRH,%
  description={Collision-Resistant Hash function . A hash function with a very low probability of two different inputs hashing to the same output}%
}%
  
\newglossaryentry{DTIME}{%
  name=DTIME,%
  description={Deterministic Time. Referring to a deterministic Turing machine's computational resource or computation time. Also TIME}%
}%
  
\newglossaryentry{Entropy}{%
  name=Entropy,%
  description={Entropy of a random variable. See Shannon entropy}%
}%
  
\newglossaryentry{FF}{%
  name=FF,%
  description={Finite Field. $\mathbb{F}$: A field that contains a finite number of elements. A field is a set with four basic operations: addition, subtraction, multiplication, and division, satisfying the rules of arithmetic. An example of a finite field is the integers modulo $\clz p$, where $\clz p$ is a prime number}%
}%
  
\newglossaryentry{FHE}{%
  name=FME,%
  description={Fully Homomorphic Encryption. An encryption scheme that enables computations to be run directly on data encrypted by the scheme without first decrypting it}%
}%
  
\newglossaryentry{HWT}{%
  name=HWT,%
  description={Hardware Token. A peripheral hardware device used to provide access to a protected or restricted electronic source}%
}%
  
\newglossaryentry{IPT}{%
  name=IPT,%
  description={Interactive Polynomial Time. The class of problems solvable by an Interactive Proof system in polynomial time}%
}%
  
\newglossaryentry{IP}{%
  name=IP,%
  description={Interactive Proof/Interactive Proof System. An abstract machine that models proof as an exchange of messages between a possibly untrustworthy Prover and an honest Verifier}%
}%
  
\newglossaryentry{MIP}{%
  name=MIP,%
  description={Multi-prover Interactive Proof. An Interactive proof system distributed across multiple provers}%
}%
  
\newglossaryentry{MT}{%
  name=MT,%
  description={Merkel Tree. A data structure in the form of a tree usually with a branching factor of 2 in which each internal node has a hash of all the information in its child nodes}%
}%
  
\newglossaryentry{NAV}{%
  name=NAV,%
  description={Non-Adaptive Verifiers. A PCP proof in which all the verifier queries are predefined}%
}%
  
\newglossaryentry{NP}{%
  name=NP,%
  description={Nondeterministic Polynomial-time. The class of decision problems that have proofs which are verifiable in polynomial time by nondeterministic Turing machines}%
}%
  
\newglossaryentry{OR}{%
  name=Oracle,%
  description={An abstract machine used in complexity theory and computability to study decision problems. Equivalent to a black box addition to a Turing machine that can solve certain problems in a single step}%
}%
  
\newglossaryentry{PCPT}{%
  name=PCPT,%
  description={Probabilistically Checkable Proof Theorem. A theorem stating that each decision problem of NP complexity can be rewritten as a probabilistically checkable proof}%
}%
  
\newglossaryentry{PCP}{%
  name=PCP,%
  description={Probabilistically Checkable Proof. A proof statement that can be checked using a randomised verification algorithm to within a high probability of accuracy by examining only a bounded number of letters of the proof}%
}%
  
\newglossaryentry{PSPACE}{%
  name=PSPACE,%
  description={Polynomial Space. The class of all decision problems that can be solved by a deterministic Turing machine using space which is polynomial in the size of the input}%
}%
  
\newglossaryentry{PTM}{%
  name=PTM,%
  description={Probabilistic Turing Machine. A nondeterministic Turing machine that uses a probability distribution to decide between alternative transitions}%
}%
  
\newglossaryentry{PVP}{%
  name={PVP},%
  description={Program Verification Problem. The problem of verifying that a computer program always achieves the intended result as given by a higher level specification}%
}%
  
\newglossaryentry{RO}{%
  name=RO,%
  description={Random or Randomised Oracle/Random Oracle model. An oracle that responds to every unique query with a random response chosen uniformly from its output domain. The response to any particular query will be the same each time the query is submitted}%
}%
  
\newglossaryentry{SH}{%
  name=SH,%
  description={Shannon entropy. The Shannon entropy of a given stochastic source is the average rate at which the source produces information. The higher the Shannon entropy, the bigger the information gained from a new value in the process}%
}%
  
\newglossaryentry{SNARG}{%
  name=SNARG,%
  description={Succinct Non-interactive Argument. A proof construction that satisfies succinctness, meaning that the proof size is asymptotically smaller than the statement size and the witness size}%
}%
  
\newglossaryentry{TM}{%
  name=TM,%
  description={Turing Machine. An abstract computing device which provides a model for reasoning about computability and its limits}%
}%
  
\newglossaryentry{VCP}{%
  name=VCP,%
  description={Verifiable Computing Problem. A computational task problem involving two agents, a relatively weak machine called the Verifier (or Client), and a more powerful Prover (or Worker). The problem involves the Verifier delegating computational tasks to the Prover. In return, the Verifier can expect the result of the task plus a proof by which it can verify the result with less computational effort than would be needed to perform the task from scratch}%
}%
  
\newglossaryentry{ZKPCP}{%
  name=ZK-PCP,%
  description={Zero-Knowledge PCP. A PCP with an additional Zero-Knowledge guarantee between the Prover and the Verifier}%
}%

\newglossaryentry{ZKSNARK}{%
  name=ZK-SNARK,%
  description={Zero-Knowledge Succinct Non-Interactive Argument of Knowledge. A Zero-Knowledge proof construction which does not involve any interaction between the Prover and Verifier}%
}%
  
\newglossaryentry{ZK}{%
  name=ZK,%
  description={Zero-Knowledge Proof. A proof construction in which the Prover can prove to another verifier that a statement is true without the Prover having to impart any information apart from the fact that the statement is true}%
}%

\newacronym{air}{AIR}{Algebraic Intermediate Representation}%
\newacronym{arm}{ARM}{Advanced RISC Machine}%
\newacronym{bcs}{BCS}{Ben-Sasson-Chiesa-Spooner}%
\newacronym{bfs}{BFS}{Byzantine Fault Tolerant NFS File System}%
\newacronym{bft}{BFT}{Byzantine Fault Tolerance}%
\newacronym{fhe}{FHE}{Fully Homomorphic Encryption}%
\newacronym{fft}{FFT}{Fast Fourier Transform}%
\newacronym{fri}{FRI}{Fast Reed-Solomon Interactive Oracle Proof of Proximity}%
\newacronym{iop}{IOP}{Interactive Oracle Protocol}%
\newacronym{iopp}{IOPP}{Interactive Oracle Protocol of Proximity}%
\newacronym{ip}{IP}{Interactive Proof}%
\newacronym{lpcp}{LPCP}{Linear Probabilistically Checkable Proof}%
\newacronym{mip}{MIP}{Multi-Prover Interactive Proof}%
\newacronym{nfs}{NFS}{Network File System}%
\newacronym{np}{NP}{Nondeterministic Polynomial time}%
\newacronym{pcp}{PCP}{Probabilistically Checkable Proof}%
\newacronym{pcpp}{PCPP}{Probabilistically Checkable Proof of Proximity}%
\newacronym{plonk}{PLONK}{Permutations over Lagrange-bases for Oecumenical Noninteractive Arguments of Knowledge}%
\newacronym{pspace}{PSPACE}{Polynomial Space}%
\newacronym{qap}{QAP}{Quadratic Arithmetic Programs}%
\newacronym{r1cs}{R1CS}{Rank-1 Constraint System}%
\newacronym{riscv}{RISC-V}{Reduced Instruction Set Computer (RISC) 5}%
\newacronym{rom}{ROM}{Random Oracle Model}%
\newacronym{rcas}{RCAS}{Reduced Cyber-Attack Surface}%
\newacronym{rs}{RS}{Reed-Solomon}%
\newacronym{rsa}{RSA}{Rivest–Shamir–Adleman}%
\newacronym{sat}{SAT}{Circuit/Boolean Satisfiability}%
\newacronym{sla}{SLA}{Service Level Agreements}%
\newacronym{snarg}{SNARG}{Succinct Non-Interactive Argument}%
\newacronym{snark}{SNARK}{Succinct Non-Interactive Argument of Knowledge}%
\newacronym{stark}{STARK}{Zero-Knowledge Scalable Transparent Argument of Knowledge}%
\newacronym{tls}{TLS}{Transport Layer Security}%
\newacronym{tpm}{TPM}{Trusted Platform Modules}%
\newacronym{uav}{UAV}{Unmanned Air Vehicle}%
\newacronym{vcp}{VCP}{Verifiable Computation Problem}%
\newacronym{zk}{ZK}{Zero Knowledge}%

\usepackage[nottoc,numbib]{tocbibind}%

\usepackage{etoolbox}%
\makeatletter%
\providecommand{\institute}[1]{
  \apptocmd{\@author}{%
    \begin{tabular}[t]{@{}c@{}}
    \end{tabular}%
    \par%
    #1%
  }{}{}%
}%
\makeatother%

\usepackage{paralist}%
\usepackage{geometry}%
\usepackage{amsmath}%
\usepackage{amsthm, amssymb}%
\usepackage[table,xcdraw]{xcolor}%
\usepackage{xcolor} %
\usepackage{circuitikz} %
\usepackage{mathtools}%
\usepackage[colorlinks=true,%
            linkcolor=gray,%
            urlcolor=gray,%
            citecolor=gray]{hyperref}%

\usepackage{booktabs} 
\usepackage{graphics} 
\usepackage{ltablex}%
\usepackage{adjustbox}%
\usepackage{multirow}%
\usepackage{cleveref}%

\usepackage{tikz}%
\usetikzlibrary{calc,fit,arrows,topaths}%

\newif\ifchanges%

\usepackage{crayola}%

\theoremstyle{definition}%
\newtheorem{definition}{Definition}[section]%

\newcommand{\field}[0]{\ensuremath{\mathbb{F}}}%
\newcommand{\pcp}[0]{\ensuremath{\mathit{\textbf{PCP}}}}%
\newcommand{\pcpp}[0]{\ensuremath{\mathit{\textbf{PCPP}}}}%
\newcommand{\np}[0]{\ensuremath{\mathit{\textbf{NP}}}}%
\newcommand{\poly}[0]{\ensuremath{\mathit{\textbf{P}}}}%
\newcommand{\ip}[0]{\ensuremath{\mathit{\textbf{IP}}}}%
\newcommand{\mip}[0]{\ensuremath{\mathit{\textbf{MIP}}}}%

\newcommand{\extractor}[1]{\ensuremath{\mathit{\epsilon_{#1}}}}%
\newcommand{\simulator}[1]{\ensuremath{\mathit{Sim_{#1}}}}%
\newcommand{\crs}[0]{\ensuremath{\mathit{crs}}}%
\newcommand{\priv}[0]{\ensuremath{\mathit{priv}}}%
\newcommand{\adversaries}[0]{\ensuremath{\mathcal{A}}}%

\newcommand{\prop}[0]{\ensuremath{\mathit{Pr}}}%
\newcommand{\sigmalanguage}[0]{\ensuremath{\Sigma^{*}}}%
\newcommand{\interaction}[1]{\ensuremath{\left\langle #1  \right\rangle}}%
\newcommand{\natpos}[0]{\ensuremath{\mathbb{N}^{+}}}%
\newcommand{\fold}[2]{\ensuremath{\mathit{Fold[#1,#2]}}}%

\newcommand{\reldistance}[2]{\ensuremath{\Delta^{*}(#1,#2)}}%

\usepackage{tikz}%
\usetikzlibrary{arrows.meta, positioning, quotes}%

\usepackage{relsize}%

\usepackage[color]{zed}%
\newcommand{\clz}{\color{ZedColor}}%

\title{State of the Art Report: Verified Computation}%
\author{%
  Jim Woodcock, \\%
  Mikkel Schmidt Andersen, %
  Diego F. Aranha, %
  Stefan Hallerstede, \\%
  Simon Thrane Hansen, %
  Nikolaj Kühne Jakobsen, %
  Tomas Kulik, \\%
  Peter Gorm Larsen, %
  Hugo Daniel Macedo, %
  Carlos Ignacio Isasa Martín, \\%
  Victor Alexander Mtsimbe Norrild \\ %
  \null\\%
  \null%
}%

\institute{Aarhus University}%

\date{31st October 2022}%

\begin{document}

\maketitle%

\clearpage%

\section*{Executive Summary}%

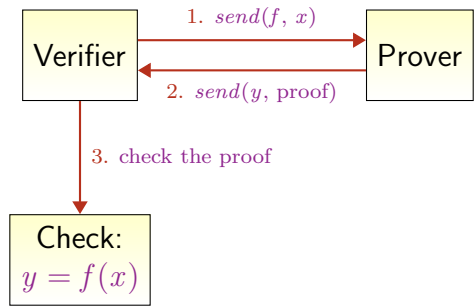
\begin{figure}[h]
  \centering%
  \begin{tikzpicture}[%
    node distance=5mm and 30mm,%
    box/.style = {%
      draw,%
      minimum height=12mm,%
      align=center,%
      top color=yellow!20,%
      bottom color=white%
    },%
    sy+/.style = {yshift= 2mm},%
    sy-/.style = {yshift=-2mm},%
    every edge quotes/.style = {align=center}%
    ]%
    \node (n1) [box]             {\textsf{Verifier}};%
    \node (n2) [box,right=of n1] {\textsf{Prover}};%
    \node (n3) [box,below=1.5cm of n1] {\textsf{Check:} \\ $\clz y=f(x)$};%
    
    \draw[thick,-Triangle,BrickRed] ([sy+] n1.east) to [above,"\hbox{\smaller\smaller 1. $\clz send(f\hbox{, }x)$}"]     ([sy+] n2.west);%
    \draw[thick,-Triangle,BrickRed] ([sy-] n2.west) to [below,"\hbox{\smaller\smaller 2. $\clz send(y\hbox{, proof})$}"] ([sy-] n1.east);%
    \draw[thick,-Triangle,BrickRed] (n1.south)      to [right,"\hbox{\smaller\smaller 3. $\clz \hbox{check the proof}$}"]    (n3.north);%
  \end{tikzpicture}
  \caption{Verifiable Computation.}\label{fig:verifiable-computation}
\end{figure}

This report describes the state of the art in verifiable computation. The problem being solved is the following:
\begin{quote}
  \textbf{The Verifiable Computation Problem (\gls{VCP})}~ Suppose we have two computing agents. The first agent is the verifier, and the second agent is the prover. The verifier wants the prover to perform a computation. The verifier sends a description of the computation to the prover. Once the prover has completed the task, the prover returns the output to the verifier. The output will contain proof. The verifier can use this proof to check if the prover computed the output correctly. The check is not required to verify the algorithm used in the computation. Instead, it is a check that the prover computed the output using the computation specified by the verifier. The effort required for the check should be much less than that required to perform the computation.  \end{quote}%
The problem is visualised in Fig.~\ref{fig:verifiable-computation}.%
This is a classic problem with many applications:
\begin{enumerate}
\item \textbf{Delegation}~ Verifiable computation can be used for delegating computation. Suppose that we have an honest prover that is efficient and runs in polynomial time (that is, the time taken is a simple function of the length of the input). Suppose also that the verifier is super-efficient and checks run in nearly linear time (that is, the time taken is directly proportional to the length of the input). The prover computes a result for the verifier. The prover then interactively proves the correctness of the result. This means the verifier can check the result's correctness nearly linearly instead of running the entire computation. The verifier does not need the computational power possessed by the prover. The prover's resources can be shared between many client verifiers.
\item \textbf{Cloud computing}~\gls{CC} Suppose we have a significant distributed computation on petabytes of data. The verifier outsources the computation and its massive dataset to a prover. The prover completes the computation and sends the results back to the verifier. The verifier wants to know that the prover executed the distributed computation correctly. The prover worries about faults that are sources of incorrect execution. These might include data corruption, communication errors, and hardware failures. The prover sends a proof that the results are free from faults.
\item \textbf{Information retrieval}~ A verifier wants to make a query on a remote database. A prover acts as the remote database server. The verifier wants assurance that the prover has performed the query correctly.
\item \textbf{Hardware supply chains}~ Hardware Trojans involve third-party threats of injecting malicious circuits into chip designs. We cannot trust hardware supply chains where this threat exists. For example, an adversarial facility might be manufacturing the hardware under contract. We want a verifier to check assurances provided by the hardware efficiently. This guarantees that the hardware is executing correctly on this input. This is a specific instance of a more general problem: verifying assertions where a third party supplies an untrusted execution substrate.
\end{enumerate}
Although we describe a three-stage protocol in Fig.~\ref{fig:verifiable-computation}, the proof might be delivered over several rounds of interaction.

The verifiable computation problem complements the program verification problem (\gls{PVP}). Verification relies on helpful redundancy. We need two descriptions of the same thing and then compare one against the other. Program verification establishes that we have expressed a given computation correctly. We make the judgement by comparing it with a higher-level specification. In the verifiable computation problem, the computation $\clz f$ is given. We are not verifying $\clz f$ against a specification. Instead, we want to know whether the execution performed by the prover is consistent with the expression of $\clz f$.

The literature surveyed in this state-of-the-art report presents the theory of probabilistic proofs. A central result in this area is the Probabilistically Checkable Proof Theorem (\gls{PCPT}). PCP has a necessary consequence. For any valid mathematical assertion, it is possible to encode the proof of that assertion. PCP shows that we can use this encoding to check the assertion's validity by inspecting only a constant number of points in the proof carried out elsewhere.

The practical consequence of PCP is in its application to the protocol in Fig.~\ref{fig:verifiable-computation}. Consider the computation $\clz f$, input $\clz x$, and supposed output $\clz y$. There is a proof and a randomised way of inspecting it that guarantees the following. The verifier will accept the proof if $\clz y=f(x)$ is correct. If $\clz y \neq f(x)$, the verifier will almost always reject the proof. The proof might need interaction between the prover and the verifier. The fact that the verifier rejects such proofs \emph{almost always} encodes an error bound. It means that, with some probability bounded in the analysis, the verifier will incorrectly view a wrong answer as correct.

The verifiable computation protocol in Fig.~\ref{fig:verifiable-computation} does not explicitly check the result $\clz y$. It does less work than that. If the verifier were to check the result $\clz y=f(x)$, then it would need to re-do the computation. That contradicts the problem statement and is not the intention.

So, PCPs allow a randomised verifier, with access to a purported proof, to probabilistically verify an input statement of the form $\clz y=f(x)$ by querying only a few proof bits. Zero-Knowledge PCPs (\gls{ZKPCP}s) enhance standard PCPs. In a zero-knowledge proof (\gls{ZK}), one party can prove to another that a given statement is true. It does this while avoiding giving any additional information apart from the fact that the statement is indeed true.

There is a large body of literature devoted to probabilistically checkable proof protocols. The original naive implementations of the PCP theory were very slow. Since then, there have been orders of magnitude improvements in performance. Early tools used low-level representations of computations. Newer tools compile programs in a high-level language for these low-level protocol entities. Some publications report efficient verifiers that could tackle real-world problems. But it appears that these systems are limited to smaller executions, mainly due to the expense of the prover. Our initial impression is that these systems are limited to special-purpose applications.

This state-of-the-art report surveys 128 papers from the literature comprising more than 4,000 pages. Other papers and books were surveyed but were omitted. The papers surveyed were overwhelmingly mathematical. We have summarised the major concepts that form the foundations for verifiable computation. The report contains two main sections. The first, larger section covers the theoretical foundations for probabilistically checkable and zero-knowledge proofs. The second section contains a description of the current practice in verifiable computation. Two further reports will cover
\begin{inparaenum}[(i)]
\item military applications of verifiable computation and 
\item a collection of technical demonstrators.
\end{inparaenum}
The first of these is intended to be read by those who want to know what applications are enabled by the current state of the art in verifiable computation. The second is for those who want to see practical tools and conduct experiments themselves.

\newpage%

\tableofcontents%

\clearpage%

\glsaddall%
\printglossary[type=\acronymtype,title=Acronyms,nonumberlist]%

\clearpage%

\glsaddall%
\printglossary[nonumberlist]%

\newpage

\section{Introduction \& Overview}

In this report, we review the state of the art in verified computation:
\begin{quote}\color{MidnightBlue}%
  \emph{How can a single computer check computations carried out by other computers with untrusted software and hardware?}
\end{quote}
We start by putting research in verified computation into context: where would it be useful?\footnote{A future deliverable in the RCAS project describes some example military applications.} A typical application is in trusted cloud services. The use of cloud computing is now widespread. The cloud is a model for enabling ubiquitous, convenient, on-demand network access to a shared pool of configurable computing resources (e.g., networks, servers, storage, applications, and services) that can be rapidly provisioned and released with minimal management effort or service provider interaction~\cite{MellG2011}. Cloud providers offer computational power and data storage facilities that significantly extend the capabilities of weaker devices. Clouds are large, complex systems that are usually provided without convincing reasons why we should trust them. In practice, they suffer from software errors, configuration issues, data corruption, hardware problems, and malicious actors. As Walfish and Blum~\cite{WalfishB2015} point out, this raises two important research questions:
\begin{inparaenum}[(i)]
\item How can we trust results computed by a third party?
\item How can we assure data integrity stored by a third party?
\end{inparaenum}
We consider several answers to these questions, suggested by Walfish and Blum~\cite{WalfishB2015}. The answers are provided by replication, trusted hardware, and remote attestation.

\paragraph{Replication} The most obvious answer to both questions is replicating computations and data over several cloud servers. For example, Canetti et al.~\cite{CanettiRR&11} discuss practical delegation of computation using multiple servers as an efficient and general solution to the problem. Their protocol guarantees the correct answer produced by a replicated, efficiently computable function. The guarantee relies on at least one honest server, but we do not know which is the honest server and which is the right answer. The protocol requires logarithmically many rounds based on any collision-resistant hash \gls{CRH} family. (Collision resistance means that it is hard to find two inputs that hash to the same output.) The protocol uses Turing Machines (\gls{TM}s) but can be adapted to other computational models. The protocol must be deterministic. This requires, in turn, deterministic versions of system calls, such as \texttt{\clz malloc()} and \texttt{\clz free()}. The construction is not based on probabilistically checkable proofs (PCP, see Sect.~\ref{sec:pcp}) or fully homomorphic encryption (FHE \gls{FHE}, see Gentry~\cite{Gentry2009}). %
Canetti et al.'s protocol, does not rely on trusted hardware, nor does it require a complex transformation of a Turing Machine program to a boolean circuit%
. The faults that must be guarded against in replicated computation are Byzantine (see Lamport et al.~\cite{LamportSP1982}): they leave imperfect evidence of the fault's occurrence. Castro and Liskov~\cite{CastroL&02} propose a novel replication algorithm, BFT, designed to handle Byzantine faults. BFT is used to implement real services, performs well, is safe in asynchronous environments like the Internet, incorporates mechanisms to defend against Byzantine-faulty clients, and can recover replicas. The recovery mechanism tolerates any number of faults over the system's lifetime, provided fewer than one-third of the replicas become faulty within a short time interval. Their Byzantine-fault-tolerant NFS file system,\footnote{Network File System (NFS) is a distributed file system protocol originally developed by Sun Microsystems in 1984, allowing a user on a client computer to access files over a computer network much like local storage is accessed.} BFS has a 24\% performance penalty compared with production implementations of the NFS protocol without replication. A major drawback of replication algorithms is that they assume failures are unrelated. This assumption is invalid for cloud services, where hardware and software platforms are often homogeneous.

\paragraph{Trusted hardware} Another solution is to use trusted hardware under the cloud provider's control. Sadeghi et al.~\cite{SadeghiSW&10} show how to combine a trusted hardware token (\gls{HWT}) with secure function evaluation to compute arbitrary functions on encrypted data where the computation leaks no information and is verifiable. Their work reduces the computation latency usually experienced by pure cryptographic solutions based on fully homomorphic and verifiable encryption.

\paragraph{Remote attestation} Yet another solution is remote attestation (see Feng~\cite{Feng2018}), where one system makes reliable statements about the software it is running on another system. The remote system can then make authorisation decisions based on that information. Many remote attestation schemes have been proposed for various computer architectures, including Intel, RISC-V, and ARM.

Walfish and Blum~\cite{WalfishB2015} give an overview of an alternative to replication, trusted hardware, and attestation: a technology where the cloud (or some other third party) returns the results of the computation with proof that the results have been computed correctly. This technology aims to make this proof inexpensive to check compared to the cost of redoing the computation. If this technology is feasible, it would not need the assumptions about faults required for replication, trusted hardware, or remote attestation. Either the proof is valid, or it is not. Walfish and Blum call this \emph{proof-based verifiable computation}. They give four insights into the practicality of this research area:%
\begin{enumerate}%
  \itemsep 0ex%
\item Researchers have already built systems for a local computer to check the correctness of a remote execution efficiently. For a comprehensive discussion of the state-of-the-art and tool implementations, see Sect.~\ref{sec:sota-and-tools}.%
\item There are diverse applications, such as computational complexity, cryptography and cryptocurrencies, secure exploitation of untrusted hardware supply chains, and trustworthy cloud computing. Additional applications, not suggested by Walfish and Blum, include computationally enhanced UAV swarms and automated reviewing systems for mathematical journals and conferences.  \item Important foundations include probabilistically checkable proofs (\gls{PCP}s)~\cite{AroraLMSS&98,AroraS&98,Sudan&09}, interactive proof systems~\cite{Babai&85,GoldwasserMR&19,LundFKN&92}, and argument systems~\cite{BrassardCC&88}, all of which have a sound and well-understood mathematical basis.  \item Engineering these theories for practical application is a new interdisciplinary research area.%
\end{enumerate}%
We describe the underlying theories in detail in later sections of this report, but for now, here is an informal motivating example due to Mordechai Rorvig and published in \emph{Quanta} magazine~\cite{Rorvig2022}:%
\begin{quote}%
  If a million computer scientists had dinner together, they’d rack up an enormous bill. And if one of them were feeling particularly thrifty and wanted to check if the bill was correct, the process would be straightforward, if tedious: they’d have to go through the bill and add everything up, one line at a time, to make sure that the sum was equal to the stated total.
    
  But in 1992, six computer scientists proved in two papers [both by Arora and his colleagues~\cite{AroraS&98,AroraLMSS&98}] that a radical shortcut was possible. There’s always a way to reformat a bill---of any length---so that it can be checked with just a few queries. More importantly, they found that this is true for any computation or even any mathematical proof since both come with their own receipt, so to speak: a record of steps that a computer or a mathematician must take.%
\end{quote}

\noindent We give a very brief overview of PCP; more details are in Sect.~\ref{sec:pcp}. A PCP for a language consists of a probabilistic polynomial-time verifier with direct access to individual bits of a bit-string. This string (which acts as an \gls{OR}) represents a proof and will be only partially accessed by the verifier. Queries to the oracle are locations on the bit-string and will be determined by the verifier’s input and coin tosses (potentially, they might be determined by answers to previous queries). The verifier must decide whether a given input belongs to the language. The verifier will always accept an input that belongs to the language, given access to the oracle (the bit string).  On the other hand, if the input does not belong to the language, then the verifier will reject with probability at least {\clz \sfrac{1}{2}}, no matter which oracle is used. One can view PCP systems in terms of interactive proof systems (\gls{IP}). The oracle string is the proof, and the queries are the messages sent by the verifier. The prover is memoryless and cannot adjust answers based on previous queries.

The PCP theorem is important. Scheideler~\cite{Scheideler2022} observes that Arora's original proof of the PCP theorem~\cite{AroraLMSS&98} is one of the most complicated proofs in the theory of computation. It has been described by Wegener as the most important result in complexity theory since Cook's theorem~\cite{Wegener2005}%
\footnote{%
  The Cook–Levin theorem, also known as Cook's theorem, proves that SAT, the Boolean satisfiability problem, is NP-complete. A problem is in \gls{NP} when the correctness of each solution can be verified in polynomial time, and solutions can be found using brute-force search (formalised using a nondeterministic Turing machine). A problem is NP-complete if it can be used to simulate every other problem for which we can verify polynomially that a solution is correct. So Cook's theorem states that SAT is in the complexity class NP and any problem in NP can be reduced in polynomial time by a deterministic Turing machine to SAT.%
} %
And by Goldreich as a culmination of impressive research rich in innovative ideas~\cite{Goldreich2008}. Boneh (quoted in Rorvig's \emph{Quanta} article~\cite{Rorvig2022}) considers it very rare that such deep algebraic tools from mathematics have made it into practice.

Both the PCP theorem and its proof have been simplified. Arora et al.'s original proof was dramatically simplified by Dinur's PCP construction~\cite{Dinur&06}. Radhakrishnan and Sudan~\cite{RadhakrishnanS2007} give an accessible commentary on Dinur's proof. Zimand~\cite{Zimand2002} presents a weaker variant of the PCP Theorem that has a correspondingly simpler proof compared to Arora et al. In Zimand's simplification, the prover has only a limited time to compute each bit of the answer, in contrast to the original prover being all-powerful. Song's account of the theoretical setting for the PCP theorem~\cite{Song} contains a simplified version of the theorem and an accompanying proof. The simplified theorem states that every NP statement with an exponentially long proof can be locally tested by looking at a constant number of bits. This is weaker than the original PCP theorem since the proofs validated by the original theorem may be much more significant. In Arora et al.'s work, the PCP verifier deals with proofs of polynomial size, whereas in Song's weaker theorem, the verifier deals with proofs of exponential size. Despite this, it is interesting that exponentially sized proofs can be verified by a constant number of queries. Ben-Sasson~\cite{BenSasson2007} presents two variants of the PCP theorem. The first achieves a nearly optimal trade-off between the amount of information read from the proof and the certainty of the proof. If the verifier is willing to tolerate an error probability of $\clz 2^{-k}$, it suffices to inspect $\clz O(k)$ bits of the proof. The second variant is very efficient in terms of the length of the proof.

\subsection{Organisation of the Report}

The report consists of the following sections.%
\begin{enumerate}%
\item The Executive Summary at the beginning of the report describes the Verifiable Computation Problem. It outlines several applications and gives a very high overview of the main foundation: the PCP theorem. A lecture given by Walfish inspires part of this description.\footnote{“Introduction and Overview of Verifiable Computation", a lecture by Prof. Michael Walfish during the Department of Computer Studies' Winter School, held by Bar-Ilan University in January 2016. See \url{www.youtube.com/watch?v=qiusq9R8Wws}.} The example of delegating computations is taken from Goldwasser et al.~\cite{GoldwasserRK&17}.
\item In Sect.\ref{sect:problem-statement}, we focus on a statement of the problem we are trying to solve with verifiable computation.%
\item The main part of this report is Sect.~\ref{sec:theory}, where we comprehensively introduce the theory of probabilistically checkable and zero-knowledge proofs. We set the scene in Sect.~\ref{sect:security-models-for-two-party-computation} by describing security models for two-party computation. We then review interactive proof systems (Sect.~\ref{sec:ips}), probabilistically checkable proofs (Sect.~\ref{sec:pcp}), and polynomial interactive oracle proofs (Sect.~\ref{sec:iop}). In Sect.~\ref{sect:computations-as-polynomial-constraints}, we discuss verifiable computation as a set of polynomial constraints that can be verified efficiently. Zero-knowledge proof protocols are used to prove that a party possesses certain information without revealing it and without any interaction between the parties proving and verifying it. We discuss such protocols in Sects~\ref{sec:snark} and~\ref{sec:stark}. Finally, in Sect.~\ref{sec:gta}, we review a game theoretic approach.%
\item The actual state of the art lies in the practice of verifiable computations embodied in usable tools. Usability includes the following considerations: the expressiveness of the language describing the computation; the efficiency of creating (and checking) the proof; and the (e.g., cryptographic) libraries on which the implementation relies. Addressing these aspects of usability poses different challenges. We review the landscape of implementations in Sect.~\ref{sec:sota-and-tools}. This is before the detailed report on the performance of practical tools on demonstrators and benchmarks.%
\item We conclude the report in Sect.~\ref{sect:conclusions} with a high-level summary of the work.%
\item The final section contains an extensive bibliography of papers and books on verifiable computation, all cited in the report.%
\end{enumerate}

\section{Problem Statement}\label{sect:problem-statement}

A verifier sends the specification of a computation $\clz P$ and input $\clz x$ to a prover. The specification of the computation might be a program text. The prover computes an output $\clz y$ and sends it back to the verifier $\clz V$ that either accepts $\clz y$ with $\clz V(y) = 1$ or rejects $y$ with $\clz V(y) = 0$. If $\clz y = P(x)$, then a correct prover should convince the verifier of this. The prover might answer some questions from the verifier or provide a certificate of correctness. If $\clz y \neq P(x)$, the verifier should reject $\clz y$ with a certain probability. The problem is to provide a protocol to carry out this procedure. The protocol is subject to three requirements.%
\begin{enumerate}%
\item The verifier must benefit from using the protocol. It might be cheaper for the verifier to follow the protocol than to directly compute $\clz P(x)$. The prover might be able to handle computations that the verifier cannot. It might have access to data inaccessible to the verifier.%
\item We do not assume that the prover follows the protocol.%
\item $\clz P$ should be general. We assume that the length of the input statically bounds the prover's running time.%
\end{enumerate}%

\paragraph{Example.}

Suppose a device $\clz D$ needs to compute the matrix product $\clz f(X) = X^T * X$ and $\clz D$ does not have the computation resources. Further, suppose another device $\clz S$ has those resources. Let us allocate the $\clz f$ computation to $\clz S$. Now we have the following scenario: $\clz D$ sends the pair $\clz (f, A)$ to $\clz S$, which computes $\clz f(A)$ and returns the result $\clz B$ to $\clz D$.

Instead of the trusted device $\clz S$ that computes $\clz f$, device $\clz D$ could also rely on an untrusted, possibly adversarial, device $\clz S'$ to compute $\clz B'$. Now, $\clz D$ needs to determine whether $\clz B'$ equals the expected $\clz B$. To ensure this, $\clz D$ requires that $\clz S'$ sends a proof $\clz\pi$ along with the result $\clz B'$ so that it now receives a tuple $\clz (B', \pi)$. On reception, device $\clz D$ verifies $\clz\pi$ is a proof of $\clz B' = f(A)$. Our main interest focuses on the two roles $\clz S'$ and $\clz D$ play in the interaction: device $\clz S'$ is tasked with producing proof and $\clz D$ with verifying it. We emphasise this by referring to $\clz S'$ as the \emph{prover} and $\clz D$ as the \emph{verifier}.

The protocol described in the preceding paragraph has a severe problem: the proof $\clz\pi$ can be prohibitively long, rendering the protocol infeasible. For the protocol to become practical, the proof needs to be shortened. Looking at only an excerpt of the proof $\clz\pi'$, we are no longer confident that $\clz B' = f(A)$. We can obtain this result only up to some probability: $\clz \exists\pi'\,\cdot\, Pr^{\pi'}[B' = f(A)] \geq c$. The value $\clz c$ is called \emph{completeness}. Of course, there is also a probability that $\clz B' \not= f(A)$. We can estimate this probability with a bound over all corresponding proofs $\clz \forall\pi'\,\cdot\,Pr^{\pi'}[B' \not= f(A)] \leq s$. The value $\clz s$ is called \emph{soundness}.

The challenge addressed by PCP is how to keep the shortened proofs $\clz \pi'$ to be considered as small as possible while maximising $\clz c$ and minimising $\clz s$.

\section{Theoretical Foundations}\label{sec:theory}

\subsection{Security models for two-party computation}\label{sect:security-models-for-two-party-computation}
Security is essential for multiparty and outsourced computation. There are many strategies to outsource computation securely. A natural approach is to rely on cryptography, which provides strong security guarantees from formal analysis grounded in time-tested hardness assumptions. Techniques not depending on cryptography can be more convenient or efficient in some contexts, but they might leave room for attacks if the incentives are large enough. We introduce several concepts: trust, verifiability, and secure communication. We discuss some of the foundational works in provable outsourced computation. Our focus is on two-party computation. Here, one entity is the client with a computational problem. The other entity (a cloud server) executes the computation and provides results to the client.

Defining security requires an adversary model against which the security properties of a system must be enforced.  Different types of adversaries capture the other threat models for two-party computation.
In general, there are three adversary models:

\begin{itemize}%
\item \emph{Semi-honest adversaries} follow the protocol specification but try to learn from a protocol execution additional information held by other parties. This adversary model is also called honest-but-curious and limits itself to confidentiality properties.%
\item \emph{Malicious adversaries} can deviate arbitrarily from the protocol specification. Protecting against these more powerful adversaries involves threats to integrity and authentication.%
\item \emph{Covert adversaries} do not necessarily follow the protocol specification but try not to be caught deviating from it. This model is a middle-ground between the previous ones and captures realistic incentive structures supporting malicious behaviour.%
\end{itemize}

\noindent The types of adversaries, non-adversarial (honest) parties, and the notion of honesty are\textbf{} further discussed in Section~\ref{sec:phm}.


\subsubsection{Trust in Multiparty Systems}
\label{sec:trust}
Trust is one of the top considerations when choosing a cloud provider for computation outsourcing. There are several definitions of trust in cloud computing~\cite{huang13,Lansing16,Pearson13}. We define trust as 
\begin{quote}
  \emph{An expectation that the cloud carries out the computation without any malicious intent, leakage of information, or purposefully returning wrong results}.
\end{quote}
Guaranteeing such a trusted relationship is complex. We present several ways trust is often established. 

A common way of determining trust is by considering the cybersecurity standard~\cite{Chung20} that the cloud provider follows. This might involve a third party's accreditation of the cloud provider based on periodic assessments. While this approach is often considered acceptable, there is no guarantee that the notion of trust holds during the execution of the client's request. In a particularly sensitive computation, such as critical utility, military, or healthcare use, cloud operators might be tempted by financial, ideological, or other incentives to temporarily violate trust. 

Another approach considers service-level agreements (SLAs)~\cite{Alhamad10} for implementing requested cloud services. This enables the client to monitor specific parameters defined within the SLA. These parameters could be the type of a machine that carries out the computation, versions of operating systems, used execution frameworks, etc. The client is informed about any changes to the agreed parameters, or the client might monitor the parameters. The cloud provider might be capable of spoofing some of the SLA parameters, or if the monitoring is random, the client might take a chance and relax the parameters. This could break the trust between the client and the cloud.

Another approach is reputation~\cite{Bilecki17} using criteria defined by the client (often based on an SLA). The client then monitors these criteria over time, analysing feedback from the client and third parties. If the criteria metric (the cloud provider's reputation) drops below some threshold, the cloud provider is no longer considered trusted. One of the challenges of this approach is the possibility of the cloud provider utilising malicious third parties to increase its reputation. A server with an honestly gained reputation may still break trust if the reward is significant enough.

These challenges have led to specialised cloud deployments. These include private clouds~\cite{Finn12}, where the client's organisation is responsible for operating the cloud platform. There may be dedicated clusters with specific security hardware, such as custom Trusted Platform Modules (TPMs)~\cite{Arthur15}. Both examples are expensive for the client, who is responsible for platform management, and in the latter case, also hardware components.

Whilst trust might be established, it cannot be fully guaranteed if the stakes for the outsourced computation are high enough. This also applies to zero trust within cloud environments~\cite{Mehraj20}. As well as cloud components not trusting external clients and devices, they may not trust each other. All access uses gateways handling authentication and authorisation. The zero-trust frameworks pose a barrier to malicious entities, including the cloud provider. But they do not guarantee the correctness of the result of the outsourced computation. This report provides an overview of techniques that provide proof of the correctness of computation and could be applied to cloud platforms without any prior establishment of trust.

\subsubsection{Verifiability in Multiparty Systems}
A potential solution to outsourced computation is verifiable computation~\cite{Fiore14}. The client dispatches the computational problem to the cloud and receives the result and proof of correctness of the result. Different approaches can be used with varying degrees of complexity. While the technical part of this report primarily discusses probabilistic and cryptographic methods. As well as receiving a result with the proof, the data and the computed problem stay hidden from the cloud. We discuss several other approaches in this section. 

The first approach is to provide \emph{verifiability by multiple executions}~\cite{Belenkiy13,Chen14}. In this approach, the same problem is distributed to various executors (cloud providers), executed, and then the results are collected and compared. This approach poses several challenges. First, it requires multiple cloud providers to be able to compute the given problem. Second, in the case of using only two cloud providers, it is not possible to determine which cloud provides the correct result (if any) without independent re-execution. Finally, the approach does not offer the ability to detect possible collusion or a state when two cloud providers provide purposefully incorrect results that are equal without utilising more than two parties (an approach based on game theory has been proposed to resolve this and is discussed later in this report (Sect.~\ref{sec:gta})).

Another approach is \emph{certifying the results by secure hardware elements}~\cite{Parno11}. In this approach, the execution is carried out with trust bootstrapped using cryptography hardware such as TPM. Any computation result is signed by keys stored within this TPM, where the client can verify this signature. In this instance, there are several challenges. The first challenge is the inability to verify the result itself (without re-execution), using only its signature. Another, perhaps more critical challenge is that trust in the physical components, such as TPMs, needs to be guaranteed, as we discussed in Sect.~\ref{sec:trust}; this is often difficult, and if there are high stakes, computing might provide a false sense of security.


Approaches based on cryptography and probability theory, such as interactive proof systems (Sect.~\ref{sec:ips}), probabilistically checkable proofs (Sect.~\ref{sec:pcp}), interactive oracle proofs (Sect.~\ref{sec:iop}), \gls{ZKSNARK} (Sect.~\ref{sec:snark}), ZK-STARK (Sect.~\ref{sec:stark}), and finally a game theory based approach (Section~\ref{sec:gta}).


\subsubsection{Honesty within Multiparty Computation}
\label{sec:phm}
Within the principles of secure multiparty computation, several considerations exist for the parties' honesty. A frequent concern is the so-called \emph{honest majority}, where most parties are assumed to have honest intentions. They do not deviate from the protocol or gather secret information. In this case, the honest majority can also be utilised to remove a dishonest party from the computation~\cite{Goldreich&87}.

In the \emph{semi-honest adversaries} case, the party does not try to deviate from the protocol specification. Instead, it tries to gather more information from the protocol execution than is allowed. This could be by studying the protocol trace or messages exchanged during protocol execution~\cite{Furukawa&19} and trying to compute additional information. Of course, there could be several semi-honest and honest parties.

In the worst-case scenario (the \emph{malicious adversaries} case), the party would willingly use any attack vector to deviate from the protocol. This deviation could be changing the inputs and outputs of the protocol, as well as aborting the protocol at any time. Security can still be achieved, especially with an honest majority. The efficiency of secure multiparty computation with malicious parties still requires improvements to reach everyday practicality~\cite{Furukawa&19, Araki&17}. The malicious party could act overtly, i.e., not trying to avoid detection or a covert (\emph{covert adversaries case}) way, making it more difficult to determine that the party is indeed malicious as the honest majority might not have enough information to support the conclusion.

An important notion is the number of corrupted parties, which may be semi-honest or malicious. The most common considerations used within the research of protocols for secure multiparty computation are $\clz t<n$, stating that the number of corrupted parties $\clz t$ could reach any number of parties within the computing environment, but less than $\clz n$: the total number of parties involved within the multiparty computation. The other common model is $\clz t < \hbox{\sfrac{n}{2}}$, where there is an honest majority. A more robust model has a two-thirds honest majority: $\clz t < \hbox{\sfrac{n}{3}}$. In many protocol cases, an honest majority is a requirement~\cite{Chida&18}. However, protocols are aimed explicitly at cases with a dishonest majority~\cite{Keller&13}. In both cases, current research is focused on the efficiency of these protocols.

\subsubsection{Secure Communication - a brief note}

Secure communication is an essential part of any data exchange. This becomes especially important in cases where the data is being sent over the network to remote parties. A typical approach is to ensure that the traffic utilises network encryption; this could be by using public key cryptography such as the TLS protocol. While the context of this report is outsourcing high-stakes computation to potentially untrusted parties, some primary considerations shall be met. It could be expected that a remote party provides a valid TLS certificate to continue outsourcing the computation. This means that the certificate shall be signed by a trusted certificate authority, limiting the remote entity's potential to fake its identity. The TLS protocol suite supported by the remote entity should also contain support for recent cypher suites and consider the use of cryptography algorithms for post-quantum TLS~\cite{Steinbach&20}. This needs to be scrutinised every time the connection is established, requiring the client to drop support for older cypher suites as they become outdated, forcing the communication to utilise only the newer ones. While there could be other protocols based on custom encryption schemes, the approaches specified in the report can be compatible with a data exchange layer based on TLS.

\subsection{Interactive Proof Systems}\label{sec:ips}

Our study of verified computation and proof systems starts with a precise definition of proof.

An alphabet $\clz \Sigma$ is typically expressed as field $\clz \field$, which defines a context for us to define objects. The set $\clz \Sigma^{*}$ defines all finite sequences of elements in $\clz \Sigma$, including the empty sequence denoted by $\clz \epsilon$.  Elements in $\clz \Sigma^{*}$ define statements and proofs.  The validity of a proof $\clz w$, for a statement $\clz x$, depends on whether $\clz (x, w)$ belongs to the relation $\clz R \subseteq \Sigma\star \times \Sigma\star$.  We define a language $\clz L_R$ to be the set of provable statements over an existing relation $\clz R$:
\begin{zed}
  L_R := \{(x, w) \in \Sigma^{*} \times \Sigma^{*} \mid (x, w) \in R\}
\end{zed}

Complexity classes divide languages into different categories depending on specific properties.  A language $\clz L$ belongs to the complexity class $\clz \np$ (the Nondeterministic Polynomial time complexity class) iff there exists a nondeterministic Turing machine (which we in our context call the verifier) $\clz V_L$, that given access to an instance $\clz x$ of length $\clz n$ and a proof $\clz \pi$ of size $\clz poly(n)$ either rejects $\clz V_L(x,\pi) = 0$ or accepts $\clz V_L(x,\pi) = 1$ the claim $\clz x \in L$ with perfect soundness and completeness.

\begin{itemize}
\item \textbf{Perfect Completeness}: $\clz x \in L \implies \exists \pi [V_L(x,\pi) = 1]$ (i.e., there exists a proof for all correct claims that will make the verifier accept).
\item \textbf{Perfect Soundness}: $\clz x \notin L \implies \forall \pi [V_L(x, \pi) = 0]$ (i.e., there exists no proof that can convince the verifier to accept a false claim).  \end{itemize}

A famous instance of $\clz \np$ is circuit/Boolean satisfiability (or simply SAT) that considers a Boolean circuit formula $\clz C$, where we want to check if $\clz w$ is a satisfying assignment $\clz C(w)=1$.  An attractive property of the Cook–Levin theorem \cite{Cook71} is that any $\clz \np$-problem can be reduced in polynomial time by a deterministic Turing machine to SAT.

Classical proofs require the verifier to read the entire proof $\clz \pi$ to be convinced of a given claim $\clz x \in L$, which in many situations becomes unattractive for a weak computational client.  A well-studied question has therefore been whether we can get away with having the verifier read fewer bits.

In the context of verifiable computation, we restrict ourselves to interactive proof systems (IP) between two parties.  Interactive proof systems were introduced in \cite{Babai&85,GoldwasserMR&19} for a function $\clz f: \Delta \to \tau$ that is a protocol that allows a probabilistic \emph{Verifier} to interact with a deterministic \emph{Prover} to determine the validity of a statement $\clz y=f(x)$ on a common input $\clz x$ based on an interaction of $\clz k$ rounds (by exchanging $\clz 2k$ messages).  The \emph{Verifier} and the \emph{Prover} interact using a sequence of messages $\clz t = (m_1, \pi_1, \ldots m_k, \pi_k)$, called the \emph{transcript}.  In the $\clz i$-th round of interaction, the Verifier challenges the Prover by sending a uniformly random message $\clz m_i$ to the Prover. The Prover replies to the Verifier with a message $\clz \pi_i$.  The Verifier is probabilistic, meaning that any challenges (messages) $\clz m_i$ sent by the Verifier may depend on some internal randomness $\clz r$ and previous messages $\clz m_0, \ldots, m_{i-1}$.  The behaviour of the Verifier can be either adaptive or non-adaptive.  The Verifier is non-adaptive (\gls{NAV}) if its queries (challenges) depend only on the Verifier’s inputs and its randomness.  An adaptive verifier selects its challenges based on previous messages.  At the end of the protocol, the \emph{Verifier} either accepts or rejects the statement $\clz y = f(x)$ based on the transcript and its internal randomness.

\begin{definition}[Interactive Proof]
  Let $\clz L \in \sigmalanguage$ be any language. We say that $\clz (P, V)$ is an interactive proof system for the language $\clz L$ if there exists a $\clz k$-message protocol between a polynomial-time verifier $\clz V$ and an unbounded prover $\clz P$ satisfying the following two properties:
  \begin{itemize}
  \item Completeness\footnote{The constants like $\clz \hbox{\sfrac{2}{3}}$ used in the definition of IP stem from the original theoretical work. They are also sufficient for proving the main PCP results. For the later, more applied work on PCP improved bounds have been derived.}: $\clz \prop[\interaction{P, V_r}(x) = 1 \mid x \in L] \geq \hbox{\sfrac{2}{3}}$
  \item Soundness against all unbounded malicious provers $\clz P^{*}$: $\clz \prop[\interaction{P^{*}, V_r}(x) = 1 \mid x \notin L] \leq \hbox{\sfrac{1}{3}}$
  \end{itemize}
\end{definition}

\begin{figure}
  \centering%
  \begin{tikzpicture}[>=angle 90,shorten >=.2em,shorten <=.2em]
    \node(P) at (-6em,0em) [rectangle,rounded corners,draw,fill=black!50,minimum width=.3em,minimum height=24em,label=above:Prover] {};%
    \node(V) at (6em,0em) [rectangle,rounded corners,draw,fill=black!50,minimum width=.3em,minimum height=24em,label=above:Verifier] {};%
    \draw[->] ($(V.west) + (0, 11em)$) -- node[yshift=.8em,rotate=9] {function $\clz f$} ($(P.east) + (0,9em)$);%
    \draw[->] ($(V.west) + (0, 9em)$) -- node[yshift=.8em,rotate=9] {input $\clz x$} ($(P.east) + (0,7em)$);%
    \draw[->,bend right=60] ($(P.west) + (0, 7em)$) to ($(P.west) + (0,4em)$);%
    \node[rectangle,draw,fill=black!10] at ($(P.west) - (6em,-5em)$) {%
      \parbox{8.2em}{%
        compile $\clz f$ into \newline%
        circuit $\clz C$ \newline%
        $\clz \downarrow$ \newline%
        compute $\clz y = C(x)$ \newline%
        keep proof \newline%
        of computation \newline%
        $\clz \downarrow$ \newline%
        encode proof%
      }%
    };%
    \draw[->,bend left=60] ($(V.east) + (0, 9em)$) to ($(V.east) + (0,6em)$);%
    \node[rectangle,draw,fill=black!10] at ($ (V.east) + (6em,7.5em)$) {%
      \parbox{8.2em}{%
        compile $\clz f$ into \newline%
        circuit $\clz C$%
      }%
    };%
    \draw[->] ($(P.east) + (0, 4em)$) -- node[yshift=.8em,rotate=351] {output $\clz y$} ($(V.west) + (0,2em)$);%
    \draw[->] ($(V.west) + (0, 1.5em)$) -- node[yshift=.8em,rotate=9] {query $\clz m_1$} ($(P.east) + (0,-0.5em)$);%
    \draw[->] ($(P.east) + (0, -1em)$) -- node[yshift=.8em,rotate=351] {response $\clz \pi_1$} ($(V.west) + (0,-3em)$);%
    \node at ($(V.west)!.5!(P.east)$)[yshift=-2.5em] {$\clz \vdots$};%
    \draw[->] ($(V.west) + (0, -4em)$) -- node[yshift=.8em,rotate=9] {query $\clz m_k$} ($(P.east) + (0,-6em)$);%
    \draw[->] ($(P.east) + (0, -6.5em)$) -- node[yshift=.8em,rotate=351] {response $\clz \pi_k$} ($(V.west) + (0,-8.5em)$);%
    \draw[->,bend left=60] ($(V.east) + (0, -8.5em)$) to ($(V.east) + (0,-11.5em)$);%
    \node[rectangle,draw,fill=black!10] at ($(V.east) + (6em,-10em)$) {%
      \parbox{8.2em}{%
        accept or reject \newline%
        proof%
      }%
    };%
  \end{tikzpicture}
  \caption{Interaction between prover and verifier}%
  \label{fig:PCP-interaction}%
\end{figure}
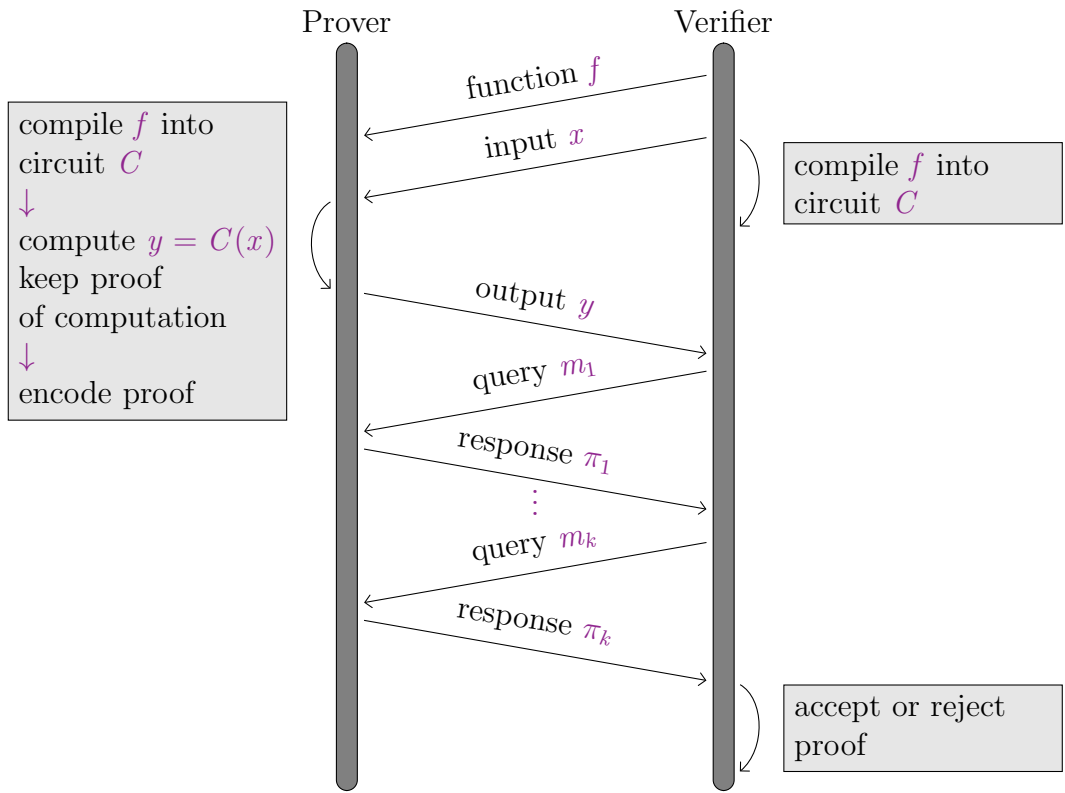

\noindent The notation $\clz \interaction{-,-}$ denotes interaction, and the subscript notation denotes the number of times the verifier draws a random bit (or \emph{coin}) to use during the verification. Figure~\ref{fig:PCP-interaction} illustrates the interaction between a prover and a verifier. Instead of computing $\clz C(x)$, a malicious prover could try to guess the proof for an invalid computation. The soundness condition expresses how hard it is to ``find'' an invalid proof that would pass the test by the verifier despite being invalid. For reasons of complexity, ``finding'' such a proof could only be achieved by resorting to randomised guessing.  An example of an interactive proof system can be found in Sect.~\ref{sec:sumcheck}.

There exist two different branches of interactive proof systems: \emph{public-coin} where the random choices (coin tosses) of the \emph{Verifier} are made public, and \emph{private-coin} where the random choices are kept secret.  Any private-coin protocol can be transformed into a public-coin protocol using the technique from \cite{GS86}.  A public-coin interactive proof is often called an Arthur–Merlin \gls{AM} game.

The complexity class $\clz \mathbf{\gls{IP}}$ denotes all languages for which an interactive proof system exists.  A fundamental measure of the efficiency of an interactive proof protocol is the round complexity $\clz \clz k$. As illustrated in Fig.~\ref{fig:PCP-interaction}, the round complexity counts the number of interactions between the prover and verifier.

It was proved in 1992 by Shamir~\cite{Shamir92} that every language in \textbf{PSPACE} has an interactive proof system ($\clz \ip = \mathbf{\gls{PSPACE}}$) that extended the work in \cite{LundFKN&92}.  Argument systems are a relaxation of interactive proof systems, where soundness is reduced to \emph{computational soundness}, meaning that soundness is only required to hold against a computationally bounded prover running in polynomial time~\cite{Wee05}.

\begin{definition}[Argument System]\label{definition:argument-system}
  An argument system $\clz (P, V)$ is defined similarly to an interactive proof system $\clz (P, V)$, with the following differences:
  \begin{itemize}
  \item The soundness condition is replaced by computational soundness: For every probabilistic polynomial time machine $\clz P^{*}$, for all sufficiently long $\clz x \notin L$, the verifier $\clz V$ rejects with probability at least {\clz \sfrac{1}{2}}.
  \end{itemize}
\end{definition}

Argument systems were introduced by Brassard et al. in 1986~\cite{BrassardCC&88} to obtain perfect zero-knowledge protocols (see Section~\ref{sec:pokzk}) for $\clz \np$.  Kilian uses the relaxation to build a constant-round protocol with low communication complexity \cite{Kilian92}.  Limiting the prover's computational power seems necessary to attain soundness of argument systems \cite{For89,GH98}.  Argument systems use cryptographic primitives and other assumptions, which a super-polynomial time prover can break. Such a prover might still be ``efficient enough'' to generate invalid proof that might pass the test.\footnote{Super-polynomial time can be ``just'' not polynomial. For instance, the Adleman–Pomerance–Rumely primality test has super-polynomial time complexity $\clz (\log n)^{O(\log\,\log\,\log\, n)}$ which gets dominated by a polynomial only for very large $\clz n$.}  Cryptographic primitives have been used to improve the situation and achieve additional desirable properties that are unattainable for interactive proofs, such as re-usability (i.e., the ability for the verifier to reuse the same “secret state” to outsource many computations on the same input), and public verifiability. Some of this work might be relevant when studying specific properties of PCP\@. A general insight here is that some properties of interest require strong assumptions on provers that limit the usefulness of protocols supporting those properties.

\subsubsection{The Sum-check Protocol}\label{sec:sumcheck}

The Sum-Check protocol\footnote{See the tutorial at \url{semiotic.ai/articles/sumcheck-tutorial/} on an excellent account of the sum-check protocol.} is an interactive proof system introduced by Lund, Fortnow, Karloff, and Nisan in~\cite{LundFKN&92}. The sum-check protocol is used in a brief account of the original proof of the PCP theorem in Subsection~\ref{sec:PCP-pt-and-insight}.  The protocol approach is very similar to how the classic television detective questions a suspect to detect whether they are lying.  The detective starts by asking for the whole story before digging into more minor details to look for contradictions.  The detective carefully selects the questions so that each communication round restricts the range of future valid answers; this means that a lying witness will eventually be caught in a contradiction. In the following, let $\clz \field$ be a finite field \gls{FF}.

The Verifier performs a similar interaction with the Prover. This is in the original setting of the sum-check protocol encoding the problem of interest as a sum-checking problem of values of a low-degree multivariate polynomial on an exponentially large hypercube.


The Prover takes as input an $\clz m$-variate polynomial $\clz g : \field^{m} \to \field$ of degree $\clz \leq d$ in each variable, where $\clz d \ll | \field |$.  The goal of Prover is to convince a verifier that:%
\begin{equation}\clz
  \beta = \sum_{x_1\in\{0,1\}} \sum_{x_2\in\{0,1\}} \ldots \sum_{x_n\in\{0,1\}} g(x_1, x_2, ..., x_n),
\end{equation}
where $\clz \beta \in \field$. The verifier has oracular access to the polynomial $\clz g$ and is given the summand $\clz \beta$.  It is the Verifier's job to determine whether $\clz \beta$ is a sum of the polynomials $\clz g(x_1, x_2, \ldots, x_n)$ for each $\clz x_i$.

The protocol proceeds in $\clz v$ rounds, where in each round $\clz i$, the Prover sends a univariate polynomial $\clz g_i$ to the Verifier.  In the first round $\clz i=1$ the Prover sends the polynomial $\clz g_{1}(X_1)$ to the Verifier, with the claim that $\clz g_{1}(X_1) = \sum_{x_2}\sum_{x_3}\dots\sum_{x_v} g(X_1, x_2, \ldots, x_v)$.  In following rounds $\clz i>1$, the Verifier selects a random value $\clz r_{i-1} \in \field$ and sends it to the Prover.  The Prover then sends a polynomial $\clz g_{i}(X_i)$ to the Verifier, with the claim that $\clz g_{i}(X_i) = \sum g(r_{1}, \ldots, r_{i-1}, X_i, x_{i+1} \ldots, x_v)$.  The Verifier checks the newest claim by considering that $\clz g_{i-1}(r_{i-1}) = g_{i}(0) + g_{i}(1)$.  In the final round the Prover sends the polynomial $\clz g_v(X_v)$ claimed to be equal to $\clz g_v(X_v) = g(r_1, \ldots, r_{v-1}, X_v)$.

The sum-check protocol has, since its introduction, been refined into different efficient proof protocols \cite{WTSTW18, ZGKPP17a, WJBSTWW17, Tha13,CMT12, GKR15}.  A significant advantage of the sum-check protocol is that implementing the prover can, in specific settings, avoid using costly operations such as the Fast Fourier Transform, which is common in other protocols.  Examples include \cite{Tha13} that implements a linear prover, or \cite{CMT12} where the prover was a streaming algorithm.


The sum-check protocol has been generalised to cover univariate polynomials \cite{Ben-Sasson&19} and tensor codes \cite{Meir13}.   

\subsubsection{The Fiat-Shamir Heuristic}
\label{subsec:fs}

The Fiat-Shamir heuristic \cite{FS86} allows the transformation of any public-coin interactive proof protocol $\clz I$ into a non-interactive, publicly verifiable protocol $\clz Q$ in the random oracle model (\gls{RO}).

The random oracle model presented in \cite{BR93, CGH98} is an idealised cryptographic model that gives all parties (the \emph{Prover} and the \emph{Verifier}) access to an entirely random function $\clz H : \{0,1\}^{k} \to \{0,1\}^{k}$ between constant sized bit strings, typically a hash function, since no such ideal function exists.  The function $\clz H$ takes an input and produces a random output chosen uniformly from the output domain.  Random oracles are, in theory, used to obtain practical, efficient, and secure protocols, while in practice are heuristically secure since no truly random hash functions can be used in practice.  Examples of this include the work in \cite{CGH98}.  They show that there are secure schemes in theory, but for which any implementation of the random oracle results in insecure schemes.  The random oracle allows the \emph{Prover} to predict the random queries (challenges) on behalf of the \emph{Verifier}.  This eliminates the need for the \emph{Verifier} to send messages to the \emph{Prover}, making the protocol inactive.

Kilian \cite{Kilian92} shows that 4-message argument systems for all of \textbf{NP} can be established by combining any PCP with Merkle-hashing based on collision-resistant hash functions.  Micali \cite{Mic00} shows that applying the Fiat-Shamir transformation to Killan's 4-message argument system yields a succinct non-interactive argument (\gls{SNARG}) in the random oracle model.  The resulting Kilian-Micali construction uses a Merkle-Tree (\gls{MT}) \cite{Merkle87} as the basis for a commitment scheme that allows sending just a single hash value (i.e., the tree’s root) as commitment.  The leaves of the tree correspond to all the Prover’s evaluation points.  Commitment schemes are discussed in the context of polynomial commitment schemes in Section~\ref{sec:polycommit}.

\subsubsection{Proof of Knowledge and Zero-knowledge}\label{sec:pokzk}

An honest prover can always convince the verifier about a true statement in a traditional interactive proof system with perfect completeness and soundness. Still, it cannot persuade it about something false. These notions can be relaxed to \emph{statistical} properties, where they hold except with negligible probability, or to \emph{computational} ones, where we admit the unlikely possibility of the prover cheating the verifier if it is computationally infeasible to do so.

A variant of this concept is \emph{proofs of knowledge}~\cite{GoldwasserMR&19}, where the prover claims to \emph{know} (or can compute) a particular piece of secret information in a way that convinces the verifier. What it means precisely for knowledge is defined as an \emph{extractor}. Since the prover cannot simply output the secret knowledge itself, a \emph{knowledge extractor} with access to the prover can extract a \emph{witness} of such knowledge~\cite{Nitulescu20zk}.

Let $\clz \clz x$ be a statement of a language $\clz \clz L$ in $\clz \clz \np$ as before, and $\clz \clz W(x)$ a set of witnesses for $\clz \clz x$ that should be accepted in the proof. Define the relation
\begin{zed}
  S = \{~ (x,w): x \in L, w \in W(x) ~\}.
\end{zed}
Proof of knowledge for relation $\clz \clz S$ is a two-party protocol between a prover and verifier in which the security notions now hold over the \emph{knowledge} of the secret value:
\begin{itemize}
\item \textbf{Knowledge Completeness}: If $\clz \clz (x,w) \in S$, then the prover $\clz \clz P$ who knows the witness $\clz \clz w$ succeeds in convincing the verifier $\clz \clz V$ of his knowledge. More formally, $\clz \clz \prop[P(x,w) \Rightarrow (V(x) = 1)] = 1$, i.e. a prover can always convince the verifier, given their interaction.
\item \textbf{Knowledge Soundness}: requires that the success probability of a knowledge extractor $\clz \clz E$ in extracting the witness, after interacting with a possibly malicious prover, must be at least as high as the success probability of the prover $\clz \clz P$ in convincing the verifier. This property guarantees that if some prover can convince the verifier using some strategy, the prover knows the secret information.%
\end{itemize}

The notion of \emph{zero-knowledge} can be seen as an additional property that an interactive proof system, an interactive argument, or a proof of knowledge can have. In this notion, there is another requirement that whatever strategy and \textit{a priori} knowledge the verifier follows or may have, respectively, it learns nothing except that the statement claimed by the prover is true. This is achieved by requiring that the interaction between the prover and verifier can be efficiently simulated without interacting with the prover, assuming the prover's claim holds. \emph{Perfect zero-knowledge} is a more robust notion of zero-knowledge that does not limit the verifier's power \cite{For89}.

We can model a cheating verifier as a polynomial-time Turing machine $\clz \clz V^*$ that gets an auxiliary input $\clz \clz \delta$ of length at most polynomial in the size of input $\clz \clz x$. The auxiliary input $\clz \clz \delta$ represents \textit{a priori} information that the verifier may have collected from previous executions of the protocol. That is, allowing the collection of this information, the verifier may cheat.

\begin{definition}[Zero-Knowledge~\cite{damgaard1998commitment}]\label{definition:zero-knowledge}
  An interactive proof or argument system $\clz \clz (P, V)$ for language $\clz \clz L$ is zero-knowledge if for every polynomial-time verifier $\clz \clz V^*$ there is a simulator $\clz \clz M$ running in expected probabilistic polynomial time such that the simulation is computationally indistinguishable (in polynomial time) from $\clz \clz (P, V)$ in input $\clz \clz x \in L$ and arbitrary $\clz \clz \delta$.%
\end{definition}

Similarly, as before, we can generalise the zero-knowledge notion as perfect (resp. statistical) zero-knowledge by replacing the requirement of computational indistinguishability with perfect indistinguishability (resp. except for negligible probability).

\paragraph{Remark.} \emph{Falsifiable} assumptions refer to the cryptographic assumptions that can be formulated in terms of an interactive game between a challenger $\clz C$ and an adversary $\clz A$ such that $\clz C$ can determine whether $\clz A$ won at the end of the game and an efficient $\clz A$ can only succeed with at most negligible probability. Intuitively, an efficient $\clz C$ can test whether an adversarial strategy breaks the assumption. That is why the majority of assumptions and constructions in cryptography are falsifiable. On the other hand, some examples of cryptographic assumptions cannot be modelled this way and are consequently non-falsifiable. Knowledge assumptions are a clear example of this phenomenon.

An example of a simple proof of knowledge protocol that also happens to be zero-knowledge is Schnorr's proof of knowledge of a discrete logarithm~\cite{schnorr1989efficient}. The protocol is defined for a cyclic group $\clz \clz \mathbb{G}$ or order $\clz \clz q$ with a generator element $\clz \clz g \in \mathbb{G}$. The prover wants to prove knowledge of $\clz \clz x = \log_g{y}$ in a group where computing $\clz \clz x$ given $\clz \clz (g,y=g^x)$ is computationally infeasible. This setting constitutes a group where \emph{computing discrete logarithms is hard}. The prover interacts with the verifier as follows:%
\begin{enumerate}
\item The prover commits to randomness $\clz \clz r$ by sending $\clz \clz t = g^r$ to the verifier.%
\item The verifier replies with a challenge $\clz \clz c$ chosen randomly.%
\item The prover receives $\clz \clz c$ and responds by sending $\clz \clz s = r+cx \bmod{q}$.%
\item The verifier accepts if $\clz \clz g^s = ty^c$.%
\end{enumerate}

The protocol can be made non-interactive using the Fiat-Shamir heuristic to hash $\clz \clz c = H(g, y, t)$ as described in Section~\ref{subsec:fs} above.  It is a valid proof of knowledge because it has a knowledge extractor that extracts $\clz x$ by interacting with the prover two times to obtain $\clz \clz s_1 = r+c_1x$ and $\clz \clz s_2 = r+c_2x$ for two distinct challenges and computing $\clz \clz x = (s_1 - s_2)/(c_1 - c_2)$.

\subsubsection{Multi-Prover Interactive Proof Systems}

The prover of an IP can be split into multiple entities (\gls{MIP}) with the restriction that these entities cannot interact while interacting with the verifier to form a multi-prover interactive proof (MIP)~\cite{BGKW88}.
The setup is reminiscent of the police procedure of isolating suspects and interrogating each of them separately~\cite{Goldreich94}.  The suspects are allowed to coordinate a strategy before they are separated. However, once they are separated, the suspects can no longer interact.  The verifier tries like an interrogator to determine if the prover’s stories are consistent with each other and the claim being asserted.

A multiple-prover proof system is more expressive than a regular IP because each prover $\clz \clz P_i$ is unaware of all messages sent to a prover $\clz \clz P_j$ where $\clz \clz j \neq i$.  It has been proved that the two-prover systems are as robust and expressive as any multi-prover interactive proof system.

Many of the ideas of MIPs have been adapted into interactive oracle proofs and probabilistically checkable proofs where a polynomial commitment scheme replaces the second prover.

\subsection{Probabilistically Checkable Proofs}\label{sec:pcp}

The first proof system introduced for $\clz \np$ in \cref{sec:ips} has perfect soundness and completeness; however, the proof size is not constant and requires the verifier to work much harder than we what we would like to be efficient and practical.  Luckily enough, research has shown that the performance of the verifier can be significantly improved if we are willing to settle for less-than-perfect soundness using a probabilistic approach where the verifier queries a random subset of the proof.

The Probabilistically Checkable Proof (PCP) theorem \cite{AroraS&98,AroraLMSS&98,Dinur&06} revolutionised the field of verifiable computation by asserting that all NP statements have a PCP, meaning that they can be written in a format that allows an efficient (poly-logarithmic) probabilistic verifier \emph{V} with oracle access to the proof $\clz \pi$ to probabilistically verify claims such as ``$\clz x \in L$'' (for an input $\clz x$ and an NP-language $\clz L$) by querying only a few bits of the proof $\clz \pi$ with soundness error $\clz \delta_{s} = \frac{1}{2}$.  The soundness error can be reduced to $\clz 2^{-\sigma}$ by running the verifier $\clz \sigma$ times.

An essential aspect of verifiable computation is the probabilistic verifier, which we will describe next.

\begin{definition}[Verifier]\label{def:verifier}
  The Verifier is a randomised Turing machine (\gls{PTM}) restricted by the following functions: $\clz l$, $\clz q$, $\clz r$, $\clz t : \natpos \to \natpos$. The Verifier has oracle access to a proof $\clz \pi$ of the statement $\clz x \in L$ with $\clz \left| x \right| = n$, where the length of the proof is restricted by $\clz \left| \pi \right| \leq l(n)$. The Verifier flips at most $\clz r(n)$ coins, queries at most $\clz q(n)$ locations of the proof to either accept ($\clz V_r(x, \pi) = 1$) or reject ($\clz V_r(x, \pi) = 0$) the statement $\clz x \in L$ in time $\clz \leq t(n)$.
\end{definition}

Given such a verifier, the complexity class of \textbf{PCP} can be constructed.

\begin{definition}[PCP class]\label{def:pcp}
  The \textbf{PCP} class is defined:
  \begin{equation}\clz
    \mathbf{PCP} 
    \begin{pmatrix}
      length &= l(n) \\
      randomness &= r(n) \\
      queries &= q(n) \\
      time &= t(n)
    \end{pmatrix},
  \end{equation}
  where $\clz l, q, r, t$ are defined as in \cref{def:verifier}.
  A language $\clz L$ belongs to the $\clz \pcp$ complexity class if there exists a 
  $\clz \mathbf{PCP}(l(n),r(n),q(n),t(n))$ with (perfect) completeness $\clz 1$ and soundness $\clz \hbox{\sfrac{1}{2}}$ for $\clz L$, where
  \begin{itemize}
  \item Perfect Completeness: $\clz x \in L \implies \exists \pi \cdot \prop[V_r(x,\pi) = 1] = 1 $
  \item Soundness: $\clz x \notin L \implies \forall \pi \cdot \prop[V_r(x, \pi) = 1] \leq \hbox{\sfrac{1}{2}}$
  \end{itemize}
  That is, an answer will always be returned. The soundness of the answer will be no worse than a half.
\end{definition}

The \textbf{PCP} class is often described as $\clz \pcp(r(n), q(n))$, which only considers the query complexity $\clz q(n)$ and the randomness $\clz r(n)$.  The verifier is described using these two functions.

Much research has studied the construction of PCPs, essentially looking for languages in the PCP class with minimal proof length, running time, randomness and query complexity while minimising the soundness error.  The most significant costs in verifiable computing are typically the verifier’s and prover’s runtime, the number of queries $\clz q(n)$, and the creation of the circuit.

\begin{definition}[PCP Theorem]\label{thm:pcp}
  The PCP theorem: $\clz \np = \pcp(O(\log\, n), O(1))$.%
\end{definition}

There have been various attempts to minimise the query complexity.  H{\aa}stad \cite{Hastad&01} showed in 1997 that the query complexity in the PCP theorem can be reduced to 3 bits.

The PCP theorem is one of the most challenging theorems in theoretical computer science and is considered one of the most important results in complexity theory \cite{Wegener2005}.  The proof of the PCP theorem is far beyond our scope here, but we point the reader to~\cite{Dinur&06}.  We will instead highlight the significant steps of the proof.

\subsubsection{Proof Techniques and Insights}\label{sec:PCP-pt-and-insight}

The early work on the PCP theorem was motivated by complexity-theoretic considerations \cite{FeigeGLSS91,FeigeGLSS96}. In contrast, later work such as \cite{Hastad&01} focused on applications where precise bounds on the performance of verifiers matter most. There has also been some work on simplifying assumptions to improve bounds and proofs of corresponding theorem variants.

The assumptions about the provers in the PCP theorem are weak, making the PCP theorem very strong. A prover can be very powerful, in particular, and have unlimited resources. Strengthening these assumptions, for instance, assuming that a prover has a (large but) limited amount of memory available, can be used to simplify proofs of corresponding variants of the PCP theorem \cite{Zimand2002}.

\paragraph{Complexity Considerations.}

In \cite{FeigeGLSS91,FeigeGLSS96}, the theorem is used to prove a result on the non-approximability of NP-complete problems. More precisely, it is shown that approximating maximal cliques in graphs is ``almost'' \textbf{NP}-complete. (The ``almost'' was later discarded by \cite{AroraS&98}.) A \emph{clique} of a graph $\clz G$ is a set of vertices $\clz C$ such that each pair of vertices $\clz n_1$, $\clz n_2$ contained in $\clz C$ is connected by an edge in $\clz G$. A \emph{maximal clique} $\clz M$ is a clique of $\clz G$ such that $\clz \left|C\right| \leq \left| M\right|$ for all cliques $\clz C$ of $\clz G$.\footnote{Cliques have many practical applications. For example, the vertices in a social network represent individuals, and the edges represent mutual acquaintance. A social network clique represents people who all know each other. Finding cliques locates groups of common friends. When Facebook makes a friend suggestion, it tries to join two cliques.}

To prove non-approximability, a variant of a $\clz \mip$ consisting of a polynomial-time Turing machine $\clz M$, that checks whether $\clz x \in L$ for some language $\clz L$, and a memoryless oracle $\clz O$. Given an input $\clz x$ (with length $\clz n$), a maximum number of random bits $\clz r(n)$ that $\clz M$ reads from $\clz x$ and a maximum number $\clz c(n)$ of communication bits exchanged between $\clz M$ and the oracle $\clz O$, one can construct a graph $\clz G_x$ whose vertices are accepting communications transcripts $\clz \langle r, q_1, a_1, \ldots q_l, a_l \rangle$ where $\clz q_i$ is a query by $\clz M$ and $\clz a_i$ is the answer of the oracle. Furthermore, $\clz r = r(n)$ and the length of $\clz \langle q_1, a_1, \ldots q_l, a_l \rangle$ must be at most $\clz c(n)$. A transcript is accepting if $\clz M$ accepts $\clz x$ with random string $\clz r$ and the history of communications $\clz \langle q_1, a_1, \ldots q_l, a_l \rangle$. The edges of $\clz G_x$ are consistent transcripts where two transcripts are consistent if identical queries in either transcript yield identical answers. The size of $\clz G_x$ is exponential in $\clz r(n) + c(n)$. Let $\clz \omega(G_x)$ denote the size of the maximal clique of $\clz G_x$. Noting that $\clz Pr_r[M^O(x, r)\hbox{\ accepts}]$ denotes the fraction of nodes for which $\clz M$ accepts and $\clz r(n)$ the number of choices made in a transcript, we get the equation $\clz max_O\, Pr_r[M^O(x, r)\hbox{\ accepts}] \cdot 2^{r(n)} = \omega(G_x)$. Using the definitions of perfect completeness and soundness given above, we have: if $\clz x \in L$, then $\clz \omega(G_x)=2^{r(n)}$; and if $\clz x \not\in L$, then $\clz \omega(G_x)\leq 2^{r(n)}/2$. This construction relates a $\clz \pcp(r(n),c(n))$ problem directly to an NP-complete clique problem of size exponential in $\clz r(n)$. It implies that solving the clique problem by approximation using $\clz M$ would also solve it. Using the exponentially-sized clique in the graph $\clz G_x$, we determine that the running time must be exponential. The PCP theorem says that --reading a few bits from the input $\clz x$ witnessed by the communications $\clz c(n)$ and tossing a few random coins $\clz c(n)$, both logarithmic in the size of the input-- a polynomial-time $\clz M$, namely $\clz n^{O(1)}$, can check whether $\clz x \in L$ for some $\clz \np$ language $\clz L$: $\clz M$ needs to be run $\clz O(\log \log n)$ times; so, it takes $\clz n^{O(log \log n)}$ times, i.e., it is ``almost'' polynomial. Iterating the execution of $\clz M$ amplifies the approximation, yielding. If approximating $\clz \omega(G_x)$ within a factor $\clz 2\log^{1-\epsilon}n$ for some $\clz \epsilon>0$ is ``almost'' polynomial, then $\clz \np$ is contained in the class of ``almost'' polynomial problems. As stated earlier, the ``almost'' can be dropped. This says that the only way an approximation can exceed the bound is if $\clz \poly = \np$.

For the applications of the PCP theorem, this insight is crucial because it implies that for the $\clz \np$-complete problems submitted to the prover, there is no deterministic polynomial algorithm that could carry out computations that could ``cheat'' the verifier. The only way is to toss a random coin and guess among an exponential number of alternatives yielding a negligible chance of cheating the verifier. For this reason, in work on the PCP theorem, complexity is often mentioned.

\paragraph{Approximation Bounds.}

The concrete approximation bounds used in \cite{FeigeGLSS91,FeigeGLSS96} are not of great interest here, only that they establish that getting a ``close'' approximation is impossible. In later publications, variants of the PCP theorem provide better bounds that are important for practical applications. We give a brief account of the original proof of the PCP theorem. Any NP-complete language can be used as a basis for the theorem. The proof uses 3-SAT. 3-SAT formulas can be arithmetised straightforwardly, which plays a central role in formulating the procedure used in the proof.

The procedure to check whether an input formula $\clz f$ of length $\clz n$ is in 3-SAT follows the five following major steps:
\begin{enumerate}
\item Arithmetise $\clz f$ with satisfying assignment (to variables) $\clz A$; the oracle knows $\clz A$, but the oracle machine does not. The following steps permit the oracle machine to evaluate whether $\clz A$ is a satisfying assignment with high probability reading only logarithmically many bits $\clz m$. The number $\clz m$ is chosen so that no smaller number satisfies $\clz 2^m=n$. So $\clz m$ is the bit width required to represent the clause indices of $\clz f$. To reduce the communication with the oracle, function $\clz f$ is extended (multilinearly) over a finite field $\clz \mathbb{F}$ such that the probability of detecting deviations by the order of the cardinality $\clz |\mathbb{F}|$ of $\clz \mathbb{F}$. To achieve a rejection probability below $\clz \frac{1}{2}$ for unsatisfying formulas later on, the size of the field must be at least $\clz 20m$. \label{it:PCPstep1}
\item Test multilinearity of $\clz A$. A function of $\clz m$ arguments is multilinear if it is linear in each argument. The assignment $\clz A$ produced in step \ref{it:PCPstep1} is multilinear, and this property is a precondition for step \ref{it:PCPstep4}. Linearity in each argument can be verified by sampling three random points of $\clz A$ from the oracle. This is sufficient to see whether $\clz A$ is ``almost'' linear for each argument with high probability because deviations of multiple segments in an argument lead to a high proportion of non-linear points. And this proportion can be detected with high probability. The ``almost'' above is described by a threshold of deviations below which the function would be accepted as linear. The number of randomly generated triples can be reduced by two-point sampling where two points are randomly generated and the remaining $\clz 20m$ points deterministically computed. Here the size of the field $\clz \mathbb{F}$ is important to achieve a rejection probability below {\clz \sfrac{1}{2}}.
\item Choose a random sequence $\clz R$ of $\clz m$ weights determining a sub-range of $\clz A$ to evaluate. This random sequence is communicated to the oracle to retrieve a portion of $\clz A$ and to evaluate $\clz f$ for $\clz A$ on the side of the oracle machine. \label{it:PCPstep3}
\item The sum-check protocol (see Subsection~\ref{sec:sumcheck}) is used to evaluate the transformed formula $\clz f$ for $\clz A$ at the sample points determined by $\clz R$ from step \ref{it:PCPstep3}. In several rounds, the oracle machine receives $\clz m$ polynomials. The size of the field $\clz \mathbb{F}$ increases the probability of discovering deviations from $\clz 0$ because the fraction of points where a multilinear function is $\clz (1-\frac{1}{\left|\mathbb{F}\right|})^{m}$. (It is $\clz 0$ for at most one value per argument. The fraction is just the ratio to the totality of points $\clz {\left|\mathbb{F}\right|}^{m}$.)  \label{it:PCPstep4}
\item The oracle machine requests the missing last three points of $\clz A$ from the sum-check protocol to complete the evaluation of $\clz f$ at $\clz A$. For steps \ref{it:PCPstep4} and \ref{it:PCPstep5} to attain the necessary probability of rejection, the field $\clz \mathbb{F}$ must have at least cardinality $\clz 100$.
  \label{it:PCPstep5}
\end{enumerate}
In the steps just listed, note that all quantities are expressed in terms of $\clz m$ and constant multiples of $\clz m$ and that $\clz m \approx \log n$. Summing up over the different logarithmic terms, we obtain the version of the PCP theorem stated in the paragraph above discussing complexity: Any language $\clz L \in \np$ is accepted by a probabilistic oracle-machine with $\clz r(n) + c(n) \leq \log n \cdot \log \log n$, and M's running time is $\clz \mathbf{DTIME}(n^{O(1)})$ (\gls{DTIME}), that is $\clz \np \subseteq \pcp(\log n \cdot \log \log n, \log n \cdot \log \log n)$ \cite{FeigeGLSS96}. An alternative proof of the PCP theorem based on \cite{Dinur&06} is discussed in \cite{RadhakrishnanS2007,Sudan&09}.

\paragraph{Improved Bounds.}

Early work on PCP provided the result that there is a constant $\clz C\geq 1$ that the optimal value $\clz \mathrm{opt}$ of specific problems cannot be approximated with a polynomial algorithm beyond $\clz C$. (Polynomial approximability is referred to as \emph{efficient} compared to approximability in general.) More formally, an algorithm $\clz C$-approximates a maximisation problem $\clz O$ if, for each input $\clz x$, it outputs a value $\clz\omega$ such that $\clz \mathrm{opt} / C \leq \omega \leq \mathrm{opt}$ (and correspondingly for minimisation problems) \cite{Hastad&01}. Concerning PCP, we are interested in inapproximability. There are differences between problems' inapproximability. No efficient algorithm can find an $\clz omege$ with the bounds. For some, $\clz C$ is a constant. For others, it is an expression depending on the input's size, and others can be approximated arbitrarily tight \cite{Moshkovitz12}. The PCP theorem also made it possible to classify these problems and get tighter bounds as the mathematical tools used in its proof evolved.

H{\aa}stad \cite{Hastad&01} presents a collection of inapproximability results. These depend very much on specific NP-complete problems. H{\aa}stad conjectures that obtaining tighter bounds for PCP depends on specialising different NP-complete problems. The improvements achieved in \cite{Hastad&01} are due to long codes \cite{BellareGS95}, their Fourier-transforms \cite{Hastad99} and joining long-code and proof verification \cite{Hastad&01}. The approximation value is calculated from the expected value of the Fourier transform of the long-code encoding of a specific approximation problem. Long codes permit encoding the problem for transmission. They lift provers' answers into a function space such that the long code of an answer $\clz x$ is a mapping $\clz A_x$ such that $\clz A_x(f) = f(x)$. It is exponential in the size of $\clz x$. Applying the Fourier-transform $\clz t$ and its inversion to $\clz A_x$ we get an expression $\clz B_x = t^{-1}(t(A_x))$ composed of sums and products with $A_x = B_x$ that permits easy calculation of expectations $\clz E[B_x]$. Testing the long code consists in choosing two random functions $\clz f_0$ and $\clz f_1$ uniformly as well as one function $\clz g$ tossing a biased random coin (with probabilities $\clz \epsilon$, $\clz 1-\epsilon$) for each value. For these the functions the product $\clz A_x(f_1)\cdot A_x(f_2)\cdot A_x(f_2)$ with $\clz f_2 = f_0 \cdot f_1 \cdot g$ is used as the acceptance condition. The expectation of the acceptance condition $\clz E[A_x(f_1)\cdot A_x(f_2)\cdot A_x(f_2)]$ can be calculated algebraically as suggested above, yielding a small chance of rejecting a correct code --because of the choice of $\clz g$-- and a logarithmic bound on the size of the size of the part of the acceptance condition $\clz A_x(f_1)\cdot A_x(f_2)\cdot A_x(f_2)$ to be evaluated. The random choice of $\clz g$ is essential to obtain the latter. The bound ensures that there can be an efficient algorithm for the verifier.

H{\aa}stad found that he can modify the algorithm for long-code testing so that it verifies Max-E3-Lin-2 and test the long code at the same time, conflating the codes to be transmitted for both so that only three values need to be requested from the prover. The completeness of the protocol is $\clz 1-\epsilon$ and its soundness $\clz \sfrac{1}{2} + \epsilon$.

If instead of a two-valued domain as in Max-E3-Lin-2 a finite Abelian group of size $\clz n$ is used, then the number of bits to be transferred is multiplied by $\clz \log n$. This corresponds to the number of bits required to encode the group's values.

Using a construction by Trevisan et al., \cite{TrevisanSSW00} called \emph{gadgets}, the bounds of Max-E3-Lin-2 can be used for a reduction in the proof of the PCP theorem \cite{Hastad&01} to derive tight bounds while reading three bits from the prover. He proves that Max-Cut cannot be $\clz (\frac{17}{16}+\epsilon)$-approximated by an efficient prover.

\subsubsection{Linear PCP}

There exist multiple flavours of PCPs. In this and the following subsections, we discuss a selection of them. We begin with linear PCP, in short, LPCP\@.

In an LPCP the proof is a linear function $\clz \lambda$, that is%
\begin{equation*}\clz
  \lambda (a * x + b * y) \ = \ a * \lambda(x) + b * \lambda(y)\ .
\end{equation*}
A language belongs to \textbf{LPCP} if there exists an $(r(n), q(n))$-LPCP-verifier $V$
\begin{itemize}
\item Perfect Completeness: $\clz x \in L \implies \exists \lambda \cdot \prop[V_r(x, \lambda) = 1] = 1 $
\item Soundness: $\clz x \notin L \implies \forall \lambda \cdot \prop[V_r(x, \lambda) = 1] \leq  \delta$,
\end{itemize}
where $\clz \lambda$ is a linear function.

LPCP are interesting because the proof can be encoded with a homomorphic encryption scheme permitting efficient implementation. The polynomial communication to the prover and constant communication to the verifier \cite{IshaiKO&07}. These properties make LPCP a practical PCP method compared to more general protocols.

Recognising such practical and useful classes of PCP appears essential for use in implementations as more general protocols have theoretical limits that cannot be overcome.

\subsubsection{PCPs of proximity}

PCPs of proximity (PCPP), also called assignment testers, relax the notion of PCPs to a problem of proximity testing~\cite{BGHSV06, DR06}, where the verifier only should test that the input is close to an element of the language instead of asserting membership.

The closeness of a language can be described using a variety of measures, but the most common one is the relative Hamming distance.

\begin{definition}[Hamming Distance]\label{def:HammingDistance}
  Let $\clz \field$ be a field, $\clz n \in \mathbb{N}^{+}$, and two string $\clz s, t \in \field^{n}$ of length $\clz n$. The Hamming distance between $\clz x$ and $\clz y$ is $\clz \Delta_n(s,t) = \left| \{ i \in \{0,.., n-1\} \mid x_i \neq y_i \} \right| $. Their relative Hamming distance is $\clz \Delta^{*}_n(s,t) = \Delta(s,t)/n$.
\end{definition}

\noindent A string $\clz s$ is $\clz \delta$-far from the string $\clz t$ if $\clz \Delta^{*}_n(s,t) > \delta$ and $\clz \delta$-close otherwise.

PCPPs refer to pair languages.  A $\clz L \subseteq \sigmalanguage \times \sigmalanguage$ is a ``pair language'' where each instance consists of pairs of strings of the form $\clz (x,w)$.  The first input $\clz x$ is the \emph{explicit} input, and the second input $\clz w$ is the \emph{implicit} input. String $\clz x$ is supposed to be in a language (characterised by proximity), and $\clz w$ is a redundant encoding of a proof used as an additional oracle by the verifier. Whereas $\clz x$ is known to the verifier in its entirety, $\clz w$ is only known ``implicitly'' and can be queried by the verifier. The idea behind providing the $\clz (x, w)$ is to be able to say that some string $\clz y$ is close to a string $\clz x \in L$ instead of determining directly whether $\clz y\in L$.  Pair languages can be defined in terms of their explicit input. Without loss of generality, we assume that the explicit input includes a specification of the length of the implicit input. Formally, the explicit input is of the form $\clz x = (x', N)$, where $\clz \left| x \right| = x'$ and $\clz N = \left|y\right|$.  For a pair language $\clz L$ and $\clz x \in \sigmalanguage$, $\clz x = (x', N)$, let $\clz L_x = \{y \in \Sigma^{N} \mid (x,y) \in L\}$.

The verifier is given the explicit input $\clz x$ and is tasked with the challenge to distinguish with high probability and with a small number of queries to the implicit input $\clz y$ to accept implicit inputs in $\clz L_x$ and reject implicit inputs that are far from being in the language.  The complexity class PCPP is defined in \cref{def:pcpp}.

\begin{definition}[PCPP]\label{def:pcpp}
  For a family of relative distance measures 
  \begin{zed}
    \Delta^{*} = \{ \Delta^{*}_N \mid \Sigma^{N} \times \Sigma^{N} \to [0, 1]\}_{N \in \natpos}
  \end{zed}
  and proximity parameter $\clz \delta \in [0, 1]$ we define the complexity class \textbf{PCPP}:
  \begin{equation}\clz
    \pcpp
    \begin{pmatrix}
      randomness &= r(n) \\
      queries &= q(n) \\
      distance &= \Delta^{*}_n
    \end{pmatrix},
  \end{equation}
  where the functions $\clz q, r$ are defined as in \cref{def:verifier}. A pair language $\clz L$ belongs to the $\clz \pcpp$ class if there exists an $\clz (r(n), q(n))$-PCPP-verifier $\clz V$ satisfying the following conditions:
  \begin{itemize}
  \item Perfect Completeness: $\clz x \in L \implies \exists \pi \cdot \prop[V_r(x,y \circ \pi) = 1] = 1 $
  \item Soundness: $\clz x \notin L \implies \forall \pi \cdot \prop[V_r(x, y \circ \pi) = 1] \leq  \frac{1}{2}$,
  \end{itemize} where $\clz \circ$ denotes composition of proofs. This is used to treat the two oracles conceptually as one.
\end{definition}

We do not define the PCPP-verifier explicitly since it is similar to the verifier in \cref{def:verifier}.

Proximity testing is often used in the context of locally testable codes.  A locally testable code is a code that a randomised Turing machine with oracular access to the supposed codeword can distinguish with a high probability between words in the code and words far from it using a small number of queries~\cite{GS06}.  An example of a locally testable code is the Reed-Solomon code~\cite{GS06}, used in the FRI protocol, described in \cref{sec:stark}.

The current state of the art in the PCPP model gives quasi-linear proofs of length $\clz n \cdot \log\, (O(1) \cdot n)$ with constant query complexity \cite{BS08,Dinur&06} and linear prover complexity $\clz O(n)$%
; the verifier complexity is $\clz \textrm{poly}\ \log n$ \cite{BGHSV06, Mie09}.


\subsection{Polynomial Interactive Oracle Proofs}\label{sec:iop}

Interactive Oracle Proof (IOP) \cite{Ben-Sasson&16,RRR16} mixes the powers of \textbf{PCP} with the interactive nature of an IP and is a generalisation of interactive PCP~\cite{KR08}.

Informally it is a $\clz k$-round interactive protocol with $\clz k$ rounds of interaction between the prover and the verifier with the twist that the verifier reads only a subset of each prover's message, referred to as oracles since the verifier has full query access.

\begin{definition}[IOP]
  An IOP system is a $\clz k$-round interactive oracle protocol defined by some relation $\clz R$ and some soundness error $\clz \delta \in [0, 1)$ and two probabilistic algorithms $\clz (P,V)$ satisfying the following properties:
  \begin{itemize}
  \item Completeness: $\clz \prop[\interaction{P, V}(x) = 1 \mid x \in R] = 1$
  \item Soundness: $\clz \prop[\interaction{P, V}(x) = 1 \mid x \notin R] \leq \delta$
  \end{itemize}
\end{definition}

This means it is an interactive protocol where the verifier runs in time sub-linearly in the total proof's length.  In each communication round, the verifier gave access/permission to a single query of the prover's message (proof).

It has been shown that any IOP can be transformed into a non-interactive argument based on the random oracle model by using the Fiat-Shamir heuristic and the transformations by Kilian and Micali~\cite{Ben-Sasson&16}.

Necessary measures of IOP protocols include the round complexity $\clz k$, along with two additional steps, namely the proof length $\clz p$, which is the total number of alphabet symbols in all of the prover's messages, and the query complexity $\clz q$.

Polynomial IOP \cite{Ben-Sasson&19} extends the IOP protocol where each message the verifier checks is a set of low-degree polynomials.  The idea of using an oracle is that rather than reading the entire list of coefficients describing the polynomial, the verifier queries these polynomials in a given point through an oracle interface, which is simulated using a polynomial commitment scheme~\cite{SZ22}.

\subsubsection{IOPs of Proximity}

IOPs of Proximity (IOPP) is a generalisation of \textbf{PCPP} to the IOP model introduced in \cite{BLNR22}.

\subsection{Computations as Polynomial Constraints}\label{sect:computations-as-polynomial-constraints}

An essential aspect of verifiable computation is the notion of computation.  We want to express a computation as a set of constraints that can be efficiently verified.  Efficiency in our context means we want to minimise the overhead associated with the interaction between the two parties and move computation to the Prover.

\subsubsection{Polynomials}

The main idea behind using polynomials is to use the mathematical foundation of the Schwartz-Zippel lemma~\cite{DL78}, which can be interpreted as follows: 
\begin{quote}
  \emph{Two different polynomials $\clz p$ and $\clz q$ of degree $\clz d$ agree on at most $\clz d$ points, formally:}
  \begin{zed}
    p \neq q \implies
    \mathit{card} (\{~ x \mid x \in \field \cdot p(x) = q(x) ~\}) \leq d,
  \end{zed}
\end{quote}
where $\clz \field$ is the finite field in which the polynomials are defined.  Consequently, the probability of finding a different polynomial $\clz Q$ that agrees with $\clz P$ in an arbitrary point $\clz x$ is $\clz d/\mathit{card}(\field)$, which is negligibly tiny if the field is large enough.  This means that polynomials are very useful for encoding the executing programs since they amplify errors made in the execution of the program, which is vital since computations are incredibly fragile---changing only a single bit in the program can change the result of the computation completely.

The probability can be further reduced by checking in more than one point, expanding the field, or reducing the degree of the polynomials.

Another fact about polynomials is that a set of $\clz n$ point-value pairs uniquely determines a polynomial of degree $\clz n-1$. This polynomial can be determined using Lagrange Interpolation~\cite{conte_elementary_2018}.

It is well known that any polynomial $\clz p(x)$ with $\clz n$ roots ($\clz r_1, \ldots, r_n$) can be described using the product of two polynomials $\clz t(x)$ and $\clz h(x)$, formally:
\begin{equation}\clz
  p(x) = t(x)*h(x),
\end{equation}
where $\clz t(x) = \prod_{i=1}^n \left(x-r_i\right)$, a polynomial with the same roots as $\clz p(x)$.  The degree of polynomials resulting from the multiplication and addition of two polynomials is well understood.  From two polynomials $\clz p(x)$ and $\clz q(x)$, where the degree of $\clz p(x)$ is $\clz n$ and the degree of $\clz q(x)$ is $\clz m$, we can create an $\clz n+m$ degree polynomial by $\clz q(x)\times p(x)$ and a polynomial of degree $\clz \max(m,n)$ $\clz p(x)+q(x)$.  Polynomials can be multiplied using the Fast Fourier Transformation in time $\clz O(n \log n)$~\cite{HH22}.

\subsubsection{Polynomial Commitment Schemes}\label{sec:polycommit}

Commitment schemes \cite{Naor1991} are used to ensure that the prover cannot modify a proof during exchanges with the verifier. The verifier must \emph{commit} to the result (of its computation) and the associated proof. Afterwards, some information is \emph{revealed} to the verifier as it accesses information received from the prover. A commitment is an encoding by the prover that fixes the prover's answer while not revealing that answer. The verifier accesses bits of the solution that are revealed without seeing the whole answer. The use of commitment schemes is widespread in cryptography applications, and they are necessary for PCP applications for the reason just stated.

Polynomial commitment schemes \cite{KateZG10} take advantage of the specific shapes of polynomials $\clz p(x)$ and their algebraic properties to permit pointwise access to the values $\clz p(i)$ of the polynomial at specific points $\clz i$ without revealing the polynomial itself.

\subsubsection{Arithmetic Circuits}\label{sec:arit_circuit}

An arithmetic circuit naturally expresses the computation of a polynomial over a field $\clz \field$, which is why they are used in verifiable computation.  Any Turing machine can be unrolled into a circuit somewhat larger than the number of steps in the computation~\cite{gennaro_quadratic_2012,PippengerF79}.  Constructing an arithmetic circuit from a program is often referred to as \emph{arithmetisation}.

An arithmetic circuit $\clz C$ of a finite field $\clz F$ is a directed graph where each vertex is either a multiplication or an addition gate with fan-in $\clz 2$, and each edge is a connection between two gates representing a wire in the circuit.  The degree of the expressed polynomial equals the number of multiplication gates in the circuit.

All gates in the same layer of a circuit $\clz C$ (viewed as a directed acyclic graph of gates) can be executed in parallel, meaning that the depth $\clz d$ of the circuit $\clz C$ represents the number of sequential steps in the program.  The size $\clz S$ of an arithmetic circuit $\clz C$ is encoded as $\clz S = | C |$: the number of gates in the circuit $\clz C$.


An arithmetic circuit is always encoded as a constraint system in the context of verifiable computation.  The most common encodings are described below.

\subsubsection{Rank 1 Constraint Systems}\label{sec:r1cs}
A constraint system defines a collection of arithmetic constraints over a set of variables to express the satisfiability of an arithmetic circuit in terms of linear constraints.

The de facto standard for these constraint systems is the Rank-1 Constraint System (R1CS), specifying an NP-complete language.  R1CS is an intermediate representation used by many tools to represent arithmetic circuits since it is easy to generate and verify.

\begin{definition}[R1CS]
  \label{def:r1cs}
  An R1CS instance is a tuple $\clz (\field, A, B, C, v, m, n)$ where $\clz A, B, C$ are $\clz m \times m$ matrices over the field $\clz \field$, $\clz v$ is a vector of maximally size $\clz m$ denoting the public inputs and outputs, and $\clz m$ is the number of variables.
\end{definition}

An R1CS instance $\clz x = (\mathbb{F}, A, B, C, v, m, n)$ is \emph{satisfiable} ($\clz SAT_{R1CS}(x,w) = 1$) if there exists a witness $\clz w$ such that $\clz (A\cdot s) + (B \cdot s) = C \cdot s$, where
$\clz w$ are the values of the private variables/wires, sometimes referred to as the witness.

We refer to the variables $\clz v$ and $\clz w$ as the public and private variables since this technique allows the \emph{Prover} to hide the private variables in the context of zero-knowledge proofs.

The matrices $\clz A$, $\clz B$ and $\clz C$ represent the structure program/circuit of interest by representing the left input of a gate, the correct input of a gate, and the output of a gate.  The vector $\clz s$ represents the values a wire/variable can take during execution on a given input.

The \emph{Verifier} generates the R1CS from the circuit; since the matrices, $\clz A$, $\clz B$, and $\clz C$ are independent of the input values, the cost of the R1CS generation can be amortised over all future computations.


\subsubsection{Quadratic Arithmetic Programs}\label{sec:qap}
Quadratic arithmetic programs (QAP) were introduced by Gennaro et al. in \cite{gennaro_quadratic_2012} to encode a program or arithmetic circuit in terms of a set of polynomials.

\begin{definition}[QAP]
  A QAP $\clz Q$ over field $\clz \field$ contains three sets of $\clz m+1$ polynomials $\clz V = \{v_{k}(x)\}$, $\clz W = \{w_{k}(x)\}$ and $\clz Y = \{y_{k}(x)\}$ for $\clz k \in \{0, \ldots, m\}$ and a target polynomial $\clz t(x)$ having the same degree and roots as the polynomial $\clz p$ described below. The constant $\clz m$ denotes the number of wires in the arithmetic circuit.
\end{definition}
The QAP $\clz Q$ computes the function $\clz F: \field^{n} \to \field^{n'}$ if there exist $\clz N = n + n'$ field elements such that $\clz (c_1, \ldots, c_N) \in \field^{N}$ is a valid assignment of $\clz F$'s inputs and outputs, if and only if there exist coefficients, i.e.\ values on internal wires, $\clz (c_{n+1}, \ldots, c_{m})$ such that $\clz t(x)$ divides $\clz p(x)$, where:
\begin{equation}\clz
  p(x) = (v_0(x) + \sum_{k=1}^{m} c_{k} * v_k(x)) * (w_0(x) + \sum_{k=1}^{m} c_{k} * w_k(x)) - (y_0(x) + \sum_{k=1}^{m} c_{k} * y_k(x))
\end{equation}
There must be a polynomial $\clz h(x)$ such that $\clz h(x)\cdot t(x)=p(x)$ for all $\clz x$.  It is a well-known fact that any polynomial $\clz p(x)$ with $\clz n$ roots ($\clz r_1, \ldots, r_n$) can be described using the product of two polynomials $\clz t(x)$ and $\clz h(x)$, formally:
$$\clz 
    p(x) = t(x) \cdot h(x),
$$ 
\noindent where $\clz t(x) = \prod_{i=1}^n \left(x-r_i\right)$ (the target polynomial).  The target polynomial is created by arbitrarily picking a root $\clz r_i$ for every multiplication gate $\clz g_i$ in the arithmetic circuit~\cite{ParnoHG016}.

\subsubsection{Algebraic Intermediate Representation}\label{sec:air}

Another approach for expressing a program as polynomial constraints is the so-called algebraic intermediate representation (AIR) introduced in \cite{BBHR18b}.  An AIR does not represent the computation as a circuit but as a trace showing how the registers and memory evolve through the execution of the program.  A $\clz T$ steps algebraic execution trace of a program $\clz P$ is represented by a $\clz \field^{w \times T}$ matrix.  A row describes the state at a given step, and a column tracks the state of a register over the entire execution.

A valid trace is determined by a set of transition constraints defining algebraic relations between two (or more) rows of the trace, along with a set of boundary constraints that enforce equality between some cells of execution and a set of constant values.

\subsection{zk-SNARKs}\label{sec:snark}

A succinct non-interactive argument (SNARG) is a triple of algorithms $\clz (G, P, V)$ where $\clz G$ is a probabilistic generator that, given the security parameter $\clz \lambda$ and a time bound $\clz T$ as inputs, outputs a reference string $\clz \crs$ and a corresponding verification state $\clz \priv$.  The honest prover $\clz P$ will, given a witness $\clz w$, produces a proof $\clz \pi \leftarrow P(\crs, y, w)$ of the statement $\clz y = (M, x, t)$ if $\clz t \leq T$.\footnote{We are using here the notation $\clz \leftarrow$ to define the variable $\clz \pi$.}  The validity of the proof $\clz \pi$ is determined by the verifier $\clz V(\priv, y, \pi)$.  The SNARG is adaptive if the prover may choose the statement after seeing $\clz \crs$. Otherwise, it is non-adaptive.

\begin{definition}[SNARG] (See \cite{Ben-SassonCGTV&13}.)
  A triple of algorithms $\clz (G,P, V)$ is a SNARG for an $\clz \np$ language $\clz L$ with a corresponding $\clz \np$ relation R, if it satisfies the following three properties:\footnote{$\lambda$ is the security parameter expressing the hardness of breaking the security scheme mentioned above. The notation $1^{\lambda}$ denotes the string $1\ldots 1$ of $\lambda$ $1$s.}
  \begin{itemize}
  \item Completeness: For all $\clz \lambda \in \natpos$ and for all $\clz (u, w) \in R$:
    \begin{equation}
      \clz%
      \prop\left[%
        \begin{array}{l|l}%
          V(\priv,u,\pi) = 1%
          &%
            \begin{array}{l}%
              (\crs,\priv) \leftarrow G(1^{\lambda}) \\%
              \pi \leftarrow P(\crs, u, w)%
            \end{array}%
        \end{array}%
      \right]%
      = 1%
    \end{equation}
  \item Computational Soundness:
    For all probabilistic polynomial-time malicious provers $\clz P^{*}$ that outputs and adaptively chosen pair $\clz (u,\pi)$ such that there is no $a$ for which $\clz (u, a)\in R$,
    \begin{equation}\clz%
      \prop\left[%
        \begin{array}{l|l}%
          \begin{array}{l}%
            V(\crs,u,\pi) = 1 \land \\%
            u \notin L%
          \end{array}%
          &%
            \begin{array}{l}%
              (\crs,\priv) \leftarrow G(1^{\lambda}) \\%
              (u,\pi) \leftarrow P^{*}(1^{\lambda}, \crs)%
            \end{array}%
        \end{array}%
      \right]%
      = negl(\lambda),%
    \end{equation}
    where $\clz negl(\cdot)$ denotes a negligible function of its input.
  \item Succinctness: The length of a proof is given by $\clz \left|\pi\right| = poly(\left|x\right| + \left|w\right|)$.
  \end{itemize}
\end{definition}
A SNARG is publicly verifiable if the private verification state $\clz \priv = \crs$, meaning that any party can verify the proofs since $\clz \crs$ is publicly known~\cite{Ben-SassonCGTV&13}.  On the other hand, a SNARG is designated unverifiable if $\clz \priv \neq \crs$ since it is only that party that knows $\clz \priv$ can verify proofs and soundness only holds if $\clz \priv$ is kept private. This definition does not address the issue of reusing the $\clz crs$ for multiple proofs.

A SNARG where the condition of computational soundness is replaced by knowledge soundness is a SNARK (succinct non-interactive arguments of knowledge)~\cite{BCCT12,BC+14}.  This means that every prover producing convincing proof must ``know'' a witness.  We define the property of knowledge soundness using a so-called knowledge extractor that can efficiently extract the witness from the prover.

\begin{definition}[SNARK]\label{def:snark}
  A triple of algorithms $\clz (G,P, V)$ is a succinct non-interactive argument of knowledge (SNARK) if $\clz (G,P, V)$ is a SNARG that comes together with a knowledge extractor $\clz \extractor{}$ as previously discussed.%
\end{definition}

A SNARK is considered a zero-knowledge SNARK if the argument does not leak information besides the statement's truth.

\begin{definition}[zk-SNARK]
  A SNARK for an $\clz \np$ language $\clz L$ defined over the relation $\clz R$ is zero-knowledge if there exists a simulator $\clz (\simulator{1}, \simulator{2})$ such that $\clz \simulator{1}$ outputs a \emph{simulated} $\clz \crs$ and a trapdoor $\clz \tau$, whereas $\clz \simulator{2}$ produces a simulated proof $\clz \pi$ given an $\clz \crs$, a statement $\clz x$ and a trapdoor $\clz \tau$. It must hold for all $\clz \lambda \in \natpos$, for all $\clz (u, w) \in R$ and for all probabilistic polynomial-time adversaries $\clz \adversaries$ (perfect zero-knowledge):
  { \clz%
    \begin{align*}
      &
        \prop
        \left[
        \begin{array}{l|l}
          \begin{array}{l}
            (x, w) \in R \land \\
            \adversaries(\pi) = 1
          \end{array} &
                        \begin{array}{l}
                          \crs \leftarrow G(1^{\lambda})\\
                          (u, w) \leftarrow \adversaries(\crs)\\
                          \pi \leftarrow P(\crs, u, w)
                        \end{array}
        \end{array}
      \right] 
      = 
      \prop
      \left[
      \begin{array}{l|l}
        \begin{array}{l}
          (x, w) \in R \land \\
          \adversaries(\pi) = 1
        \end{array} &
                      \begin{array}{l}
                        (\crs, \tau) \leftarrow \simulator{1}(1^{\lambda})\\
                        (u, w) \leftarrow \adversaries(\crs)\\
                        \pi \leftarrow \simulator{2}(\crs, u, \tau)
                      \end{array}
      \end{array}
      \right]
    \end{align*}
  }%
  The simulator $\clz (\simulator{1}, \simulator{2})$, using some trapdoor information $\clz\tau$, can find a proof without consulting the witness $\clz w$, i.e., it does not require any knowledge about the witness.
\end{definition}
SNARKs have been extensively studied and are used in many different contexts and applications.  There are many different variants of SNARKs. Nevertheless, they all arise from an underlying characterisation of the complexity class $\clz \np$ as a specification of an NP-complete problem with specific advantageous properties.  A popular starting point for many of the most influential SNARKs~\cite{gennaro_quadratic_2012, Lip13, DFGK14, Groth16, GMNO18} has been the NP-complete decision problem circuit satisfiability (Circ-SAT).  Many of the mentioned SNARKs for circuit satisfiability are based on the characterisation of quadratic span programs and quadratic arithmetic programs introduced by Gennaro et al.\ in \cite{gennaro_quadratic_2012}.

\begin{definition}[Circuit Sat]
  The Circ-SAT problem is the $\clz \np$-complete decision problem of determining whether a given circuit has an assignment of its inputs that makes the output true. A circuit $\clz C$ is a function $\clz C : \{0, 1\}^\ell \rightarrow \{0, 1\}$. The decision problem consists in finding a vector of constants $\clz (a_1, \ldots, a_\ell)$ such that $\clz C(a_1, \ldots, a_\ell) = 1$ if it exists.%
\end{definition}

\subsubsection{Examples of constructions}

Designing zk-SNARKs that produce short constant-sized proofs and efficient prover/verifier algorithms is a hot research topic. We provide an overview of some recent examples below. We note that all these constructions are based on specific results from group theory that make the computation of discrete logarithms hard. Consequently, the algorithms cannot be quantum-safe or offer long-term security against quantum adversaries. In practice, these groups are typically instantiated with cryptographically secure \emph{elliptic curves}~\cite{cryptoeprint:2022/586}.

\paragraph{Groth16.} It was the first practical zk-SNARK based on cryptographic pairings~\cite{Groth16}, and thus is used as a reference point for comparison and benchmarking. The circuits are represented as a QAP, and proofs are quite compact (only three group elements), much shorter than~\cite{ParnoHG016}. The main drawback with this construction was relying on a public reference string used by the prover and verifier that was created using a one-time trusted setup. The setup was circuit-specific, so it had to be repeated for any new circuit. If the setup is performed maliciously, soundness is broken, and undetectable fake proofs may be generated. A proposed way to mitigate this problem was to conduct complex multi-party computation for the setup. The most famous practical deployment of this zk-SNARK construction was the ZCash cryptocurrency~\cite{Ben-SassonCG0MTV14}.

\paragraph{Bulletproofs.} It is a proof system introduced by Bünz et al.~\cite{Bunz&18} that does not require a trusted setup or cryptographic pairings to be efficient, thus relying on weaker assumptions. However, it is not a zk-SNARK \emph{per se} since proof sizes scale logarithmically and verification time scales linearly with the size of the circuit, even when batching is applied. Bulletproofs is still quite practical and competitive for simpler circuits, for example, (batched) range proofs to prove that integers lie in a certain interval. For this reason, it is a popular option for proof systems implementing confidential transactions in cryptocurrencies.

\paragraph{Sonic.} It is an early and quite practical example of a polynomial IOP~\cite{MallerBKM19}, as described in Section~\ref{sec:iop}. It represents the circuit as three vectors of left, right and output wires, for which the consistency for multiplication and addition gates is reduced to one large polynomial equation. This equation is embedded into the constant term of a bivariate polynomial. The verifier checks that the bivariate polynomial is computed by the prover from the witness and circuit such that the constant term is cancelled out.  Sonic proofs are small, the prover is efficient, and verification is fast when many proofs for the same circuit are verified simultaneously. This fits many applications, for example, checking a batch of financial transactions for consistency, where Sonic remains state-of-the-art. Sonic also enjoys an optional \emph{helper} mode, in which a third party commits to a batch of proofs and provides a short and easy-to-verify proof of the correct calculation, making the final verification more efficient. Because of its custom constraint system, Sonic implementations typically offer an adaptor for circuits in R1CS.

\paragraph{Gates.} A polynomial division check ensures that the prover knows satisfying inputs and outputs for each gate (witness), and a permutation argument provides correct wiring between the gate outputs and inputs. The constraint system is non-standard and richer than all previous constructions, supporting different operations and optimizations, which perform at some cost in complexity in the PLONK compiler. PLONK offers shorter proofs and proving times in comparison to Sonic.

\paragraph{Marlin.} It is an improvement over Sonic with much faster prover and verification times for individual proofs~\cite{ChiesaHMMVW20}. While Sonic remains state-of-the-art for batched settings, Marlin shines when batching is impossible.  Compared to PLONK, the proving cost and proof size are similar, with the advantage that Marling uses an R1CS constraint system that is much more standard and widely supported by existing tooling.

\subsection{zk-STARKs}\label{sec:stark}

The most crucial difference between a zero-knowledge scalable transparent argument of knowledge (zk-STARK) and a zk-SNARK is the transparency of the zk-STARK approach. As a consequence of this transparency, it is also post-quantum secure. The zk-STARK approach was introduced by Ben-Sasson et al.~\cite{BBHR18b} and is based on the FRI protocol described below.

\subsubsection{Fast Reed-Solomon IOP of Proximity}\label{sec:fri}

Fast Reed-Solomon Interactive Oracle Proof of Proximity (FRI) was introduced by Ben-Sasson et al.~\cite{BBHR18a} and gradually improved concerning the soundness bound of FRI~\cite{BKS18, BGKS20, BCIKS20}.  The protocol is an IOPP working over a pair language of Reed-Solomon (RS) codes.

\begin{definition}[Reed-Solomon Codes]
  The evaluation of a polynomial $\clz P(z) = \sum^{d}_{i=0} a_iz^i$ over $\clz S \subseteq \field$, with $\clz \field$ a finite field and $\clz \left|S\right| = n$, is the function $\clz p : S \to \field$ defined by the function $\clz p(s)=P(s)$ for all $\clz s \in S$.
  The code $\clz RS[\field, S, \rho]$ is the space of functions $\clz f : S \to \field$ that are evaluations of polynomials of degree $\clz d < \rho \cdot n$.
  Formally,
  \begin{equation}\clz
    RS(\field, S, \rho) = \{f : S \to \field \mid deg(f) < \rho\cdot\left|S\right|\}
  \end{equation}
\end{definition}
The pair language PAIR-RS is defined as follows.  The direct input is the definition of the RS code $\clz (\field, S, \rho)$, where $\clz \field$ is a finite field, $\clz S \subseteq \field$, and $\clz \rho \in [0,1]$ is the code rate. The code rate describes the amount of helpful information in a string: if a string has length $\clz n$ (as above) and the code rate $\clz\rho$ is a fraction of $\clz n$, then the string contains $\rho\cdot n$ useful bits and $\clz n - \rho\cdot n$ redundant bits.  The verifier's task is to distinguish with high probability and with a small number of queries to the implicit function $\clz f: S \to \field$, between the case that $\clz f \in RS[F, S, \rho]$ and the case that $\clz f$ is $\clz \delta$-far from (all members of) $\clz RS[F, S, \rho]$ in relative Hamming distance.

We define the distance of an element $\clz e \in \Sigma^{N}$ and a set $\clz S \subseteq \Sigma^{N}$ to be
\begin{equation}\clz
  \reldistance{e}{S} = 
  \begin{cases*}
    \min_{s \in S} \reldistance{e}{s}  & if $\clz S \neq \emptyset$ \\
    1             & $\clz S = \emptyset$
  \end{cases*}
\end{equation}

Checking if $\clz f \in RS[F, S, \rho]$ is essentially a variant of low-degree testing, which can be naively performed using $\clz d+2$ queries ($\clz d+1$ queries to recover the polynomial and $\clz 1$ query to verify), which is far worse than the logarithmic query complexity of the FRI protocol.

We will not describe the FRI protocol in detail but briefly overview its essential components.  The FRI protocol is inspired by the Fast Fourier Transform (FFT), from where it inherits the word ``fast'' in its name and many properties.  The FRI protocol uses the proof composition introduced for PCPs in~\cite{AroraS&98} adapted for the use with RS to reduce the problem of proximity testing of a code word $\clz f^{i}$ to the RS code $\clz RS[F, S, \rho]$ to the problem of testing the proximity of a linear combination of two code words $\clz f^{i+1}$ to $\clz RS[F, S', \rho]$ where $\clz \left|S'\right| = \hbox{\sfrac{1}{2}}\cdot\left|S\right|$, which effectively is a problem of half size involving low-degree testing of a polynomial of degree $\clz < \rho\cdot\left|S\right|\cdot\hbox{\sfrac{1}{2}}$.  Proof composition refers to encoding proofs that permit random access to different parts that can be composed instead of reading (and re-reading) larger parts of a proof. The PCP approach was adapted to PCPPs in~\cite{BGHSV06, DR06, BS08}.  An essential property of this proof composition is distance preservation meaning that if $\clz f^{i}$ is $\clz \delta^{i}$-far from all code words, $\clz \reldistance{f^{i}}{RS[F, S, \rho]} \geq \delta^{i}$, then it must be the case that $\clz f^{i+1}$ is $\clz \delta^{i+1}$-far from all code words, $\clz \reldistance{f^{i+1}}{RS[F, S, \rho]} \geq \delta^{i+1}$, where the reduction in distance between rounds is negligible.  The FRI protocol improves significantly on distance preservation compared to previous constructions for PCPPs and IOPPs based on the bivariate testing Theorem of Polischuk and Spielman~\cite{PS94}, which suffers a constant multiplicative loss in the distance per round of proof composition $\clz (\delta^{i+1} \leq \delta^{i}/2)$.

Proof composition of the RS code $\clz RS[F, S, \rho]$ to $\clz RS[F, S', \rho]$ is called \emph{folding} and can maximally be repeated until the degree of the polynomial is $\clz 0$.  The FRI protocol relies on the following three observations~\cite{BLNR22}.

\begin{enumerate}
\item Decomposition of polynomials: Any polynomial $\clz f$ where $\clz deg(f)=d$ can be split into two polynomials $\clz f_e$ and $\clz f_o$ where $\clz deg(f_e,f_o) < \frac{d}{2}$ such that:
  \begin{equation}\clz
    f(x) = f_e(x^2) + x*f_o(x^2)
  \end{equation}
  This property is also used for using FFT to multiply polynomials where $\clz f_e(x) = \frac{f(x)+f(-x)}{2}$ represent the even coefficients and $\clz f_o(x) = \frac{f(x)-f(-x)}{2x}$ the odd coefficients. While $\clz f$ is a polynomial over the domain $\clz S \subseteq \field$, the functions $\clz f_o$ and $\clz f_e$ have domain $\clz S^2 = \{x^2 \mid x \in S\}$, where $\clz S^2 \subseteq S$ since $\clz x^2 = (-x)^2$.
  
\item Randomised folding/hashing: Let $\clz S$ be a multiplicative group of order $\clz 2^r$ with the generator $\clz \omega$. Let $\clz f$ be a polynomial over $\clz S$ with degree $\clz d$. Then, define $\clz S' = \langle \omega^2 \rangle = \{x^2 | x \in S\}$ such that $\clz \lvert S' \rvert ~=~ \lvert S \rvert/2$. Let $\clz \pi : \field \to \field$ be the map defined such that $\clz \pi(S) = S'$. The structure of the evaluation domain allows the reduction of the proximity problem to one of half the size at each round of interaction.
  
  Based on the decomposition property of polynomials, a folding operator $\clz \fold{f}{z} : \field^{S} \to \field^{S'}$ is defined. The operator maps for any $\clz \alpha \in \field$ a code word in $\clz RS[\field, S, \rho]$ to a code word in $\clz RS[\field, S', \rho]$. The folding is defined as follows:
  \begin{equation}\clz
    \fold{f(x)}{\alpha} = g(x^2) + z*h(x^2)
  \end{equation}
  where $\clz deg(g) < d/2$ and $\clz deg(h) < d/2$ if $\clz deg(f) = d$. Moreover we have the relationship $\clz f(x) = g(x^2) + x*h(x^2)$.
  
\item Folding is Distance preserving: Except with small probability over $\clz \alpha$, the folding of a polynomial $\clz f$ is distance preserving. Formally,
  \begin{equation}\clz
    \reldistance{f}{RS[\field, S, \rho]} \geq \delta \implies \reldistance{\fold{f}{z}}{RS[\field, S', \rho]} \geq (1-o(1))\delta
  \end{equation}
\end{enumerate}
FRI follows a simple repeating pattern of $\clz r$ rounds of interaction divided into two phases Commit and Query.  The verifier sends in each round a random challenge $\clz \alpha \in \field$ to detect if $\clz f^{i} \in RS[\field, S^{i},\rho]$.  The prover answers by committing to a new code word/oracle function $\clz f^{i+1} : S^{i+1} \to \field$, which it claims to be equal to $\clz \fold{f^{i}}{z}: S^{i+1} \to \field$.  Each round reduces the problem by half, eventually leading to a function $\clz f^{r}$ evaluated over a small enough evaluation domain. This induces a sequence of Reed-Solomon codes of strictly decreasing length.  Nevertheless, the code $\clz \rho$ and the relative minimum distance remain unchanged.  Finally, the verifier ensures that $\clz f^{r}$ belongs to the last RS code.

The verifier checks the consistency between related code words $\clz f^{i}$ and $\clz f^{i+1}$.  The verifier queries $\clz f^{i+1}$ in a single point $\clz z \in S^{i+1}$ and selects two corresponding points $\clz z_{+},z_{-} \in S^{i}$ such that $\clz \pi(z_{+}) = \pi(z_{-}) = z$.  The consistency check interpolates between the two points $\clz (z_{-}, f^{i}(z_{-}))$ and $\clz (z_{+}, f^{i}(z_{+}))$ to create the maximally one degree polynomial $\clz p(x)$, that should pass through all three points $\clz (z_{-}, f^{i}(z_{-}))$, $\clz (z_{+}, f^{i}(z_{+}))$ and $\clz (z, f^{i+1}(z))$.  Efficient implementations save query complexity and boost soundness by reusing the oracle queries to test the consistency of $\clz f^{i-1}$ vs. $\clz f^{i}$ and consistency of $\clz f^{i}$ vs. $\clz f^{i+1}$.  Efficient implementations use the Fiat-Shamir transform to achieve a non-interactive protocol.

\subsection{Game theory approach}\label{sec:gta}

In the search for lower-cost approaches to verifiable computation, an approach based on game theory and smart contracts has been proposed. Lowering costs is based on limiting (or avoiding) heavy cryptographic protocols. In cases that utilise this approach, a client outsources the computation to multiple parties while also creating distrust between the computing parties.

An approach that considered outsourcing computation to two execution parties while creating distrust between them by use of smart contracts~\cite{Zheng&20} have shown promising results~\cite{Dong&17}. In general terms, the approach avoids cryptography to remove the possibility of collusion between computing entities. At the same time, it focuses on utilising economic means to create distrust between the computing entities, which leads to the prevention of collusion. Collusion prevention is essential in this approach as it is based on \emph{replication}, meaning that it outsources the same problem to multiple computing entities and compares results from them. In case of collusion, several computing parties can collude to provide a wrong result, which could be undetected by the client that ordered the computation. The approach works by making collusion the least favourable option for computing entities.

The computing entities are asked to sign two contracts, a so-called \emph{Prisoner contract} and a \emph{Traitor contract}. The prisoner contract is created between the client ordering the computation (paying a fee $\clz cl_c$) and two execution parties, where the execution parties take a fee $\clz c$ to carry out the computation. The incentive is based on a deposit that each cloud is asked to pay in advance that will be refunded in case of correct computation. In addition, the contract specifies cases where an honest party receives the deposit confiscated from the cheating party. This leads to a higher payoff for honesty than collusion and incentivises one party to try to get the other party to cheat while remaining honest. This causes instability for collusion since both parties are aware of the incentive structure. The parties further agree on specific deadlines $\clz T_i$ to enforce timeliness and ensure that the parties move forward in execution. If the dispute arises between the two parties, it needs to be settled by a trusted third party $\clz \cal{T}$ that takes a fee $\clz c(\cal{T})$. The computation problem $\clz f()$, including the input $\clz x$ provided by the client to the executing parties, could or could not be hidden depending on the cost sensitivity of the ordering client. Each of the executing entities is asked to pay a deposit $\clz d$ where the game analysis has demonstrated that given that $\clz d > c + c(\cal{T})$ the game reaches an equilibrium where both parties execute $\clz f(x)$ with probability $\clz 1$ and return a correct result while respecting $\clz T_i$.

Before exploring the traitor contract, it is essential to explore a situation where executing entities create a so-called \emph{Colluder's contract}. The Colluder's contract is introduced to create a position where the executing parties could make enforceable promises to each other and be able to subvert the prisoner contract, hence making collusion an attractive option again. This contract introduces a bribe $\clz \cal{B}$ paid by the entity that has initiated the collusion and a collusion deposit $\clz d(c)$ that both parties pay and the party that does not execute the collusion loses. The contract needs to be taken into effect before the solutions from the executing parties are provided back to the client. The game analysis has shown that upon satisfying conditions such as $\clz {\cal B} < c \land cl_c - {\cal B} > cl_c - c$ then the collusion provides the highest payoff and hence it is often utilised.

Finally, to avoid the issue of colluding parties, the traitor contract provides a solution. This contract allows an execution entity to be exempt from paying a penalty for collusion specified within the prisoner contract, providing an incentive for reporting the collusion. If collusion is reported it is up to the $\clz \cal{T}$ to determine who has cheated in computation. This contract allows for one party to betray the other party secretly, and the betrayal is risk-free. Hence, this contract destabilises the potential for collusion by creating distrust between execution entities. The game analysis based on including all three contracts has shown that execution entities eventually behave honestly. Most of the time, no execution party wants to initiate collusion.

A variation on this approach could be utilised in a setting with more than two execution parties that can achieve strict Nash equilibrium given a correct set of incentives~\cite{Alptekin&17}. The client first \emph{randomly} chooses different execution entities and then assigns a job to them. Once the job is completed, the client checks if the provided answers match. If this is the case, all execution entities are rewarded. If the answers do not match, the client dispatches the same problem to another randomly selected population of execution entities and repeats the step until a matching result is obtained from all of the execution entities. This is then further utilised to punish, i.e. fine the execution entities that returned wrong results. To ensure that the execution entity can pay a fine in case of a wrong result, the client first checks all of the execution entities it is about to hire for funds available to ensure this payment. One issue is in the case of colluding execution entities, where all entities collude to provide the same answer to the client. To do that, however, all the colluders would need to keep some budget for a fine; otherwise, they would not be hired. This means they would need to behave honestly, at least sometimes, to keep their budget from becoming negative. This led to a bound of the maximum damage colluders can cause to the client could be evaluated as extra work defined as $\clz rgm/(r + f)$, where $\clz r$ is the reward for the execution entities, $\clz g$ is the amount of colluding entities, $\clz m$ is the total amount of execution entities, and $\clz f$ is fine for wrong computation.

While the approaches based on incentives and game theory could assure that the computation is carried out correctly, it often requires a certain level of re-execution (either local or with another set of remote execution entities). Hence, it is not clear of overhead. Another issue that this approach does not solve is hiding the function and the inputs being computed. This still requires the use of cryptographic methods described previously.



\section{Practice in Verifiable Computations}
\label{sec:sota-and-tools}


The ultimate goal of the research in verifiable computation is to enable computation outsourcing to third parties and obtain the computation results at an asymptotically lower computational cost than performing the actual computation locally. Nevertheless, given the additional overhead the proof system adds, achieving this goal is a challenge to the theoretical foundations because devising an efficient proof system implementation is far from trivial, to the point that authors have characterised existing implementations as ``near practical'' \cite{WalfishB2015}. In this section, we report on our findings after performing experiments to study the feasibility of using verifiable computations toolchains as a component off-the-shelf.  More precisely, given a user-defined program, is it possible to use the toolchain to produce a prover, a verifier, and use a verifiable computation proof system without hacking its implementation?

\subsection{Implementations: The Bottleneck}
A proof system for verifiable computation translates the original computation (obtain the result $\clz y$ from running $\clz P$ on input $\clz x$) into (abstractly\footnote{Given the diversity of proof protocols there may be more or fewer steps involved.}) three computation steps:
\begin{enumerate}
\item The verifier setup: producing a specification $\clz \Sigma$.
\item The prover computation: producing $\clz y' = \Sigma(x)$ and a certificate $\clz \pi$.
\item The verifier check: checking the logical assertion: $\clz \pi \implies y = y'$.
\end{enumerate} 
These steps have varying complexities depending on each approach but are still not generally applicable. According to the latest accounts \cite{Wahby&21}, the prover computation (step 2) adds an overhead of at least $\clz 10^7 * T$ (where $\clz T$ is the time to compute $\clz y$), the verifier adds at least a linear computation overhead on the input and output to check the proof $\clz \pi$ (step 3) and also incurs a cost proportional to running the program when transforming the original program $\clz P$ (step 1) into a specification $\clz \Sigma$, a suitable description of $\clz P$ in the computational model supported by the prover.  Step 1 is necessary, given that developers very seldom specify their programs using any of the computational models required by the proof systems, e.g., arithmetic circuits or polynomials. Usually, a program is first specified in a high-level programming languages like Java, C, or Python.

The implementation systems available as mature research contributions, and the ones we cover in our study, consist of toolchains transforming the program $\clz P$ into prover and verifier as toolchain elements that work as a ``compiler'', which typically accepts some subset or superset of C. The compilation process may be architecturally divided into front-end and back-end components.
The front end compiles the program into one of the above-described models suitable for probabilistic checking. 
The technique for turning a computation into a constraint system involving native operations over a finite field is called
\emph{arithmetisation}.
The back end is an interactive proof or argument system that is applied to verify the correctness of the statement.

\subsection{Surveying Implementations: A Selection of Tools}

We covered tools with a publicly available version-control repository. We selected an assortment that fits our project goal: to assess the usage of research projects as off-the-shelf in projects where verified computation would run in practice. In the following, we introduce the particulars of each tool we evaluate.

\paragraph{Pequin.} This tool is the amalgamation of several implementations of the theoretical results in~\cite{SettyMBW&12,SettyVPBBW&12,VuSBW&13,BraunFRSBW&13,WahbySRBW&15} produced over several years of research by the Pepper project, which began with the original Pepper tool. It is a refactored version of the Pepper library made more readable and easily accessible, featuring easy-to-use Docker files and tutorials. Pequin has the option to run with the implementation of five different implementations: \textbf{Ginger}, \textbf{Zataar}~\cite{VuSBW&13}, \textbf{GGPR12}, \textbf{Pinocchio}, and \textbf{PinocchioZK}. These protocols can be set from inside the utility header file. This allows the user of the tool to quickly switch between the different protocols based on the requirements of the tool. It is very intuitive to work with Pequin since writing proofs is done via a simple \emph{compute} function that takes some inputs and outputs that the user can define. Pequin is being maintained and updated. It has ongoing issues being handled, making it an active and maturing tool.

\paragraph{Cairo.} This is a Turing-complete, production-ready language for creating zk-STARKs~\cite{Goldberg&21}. It was written and maintained by \textbf{StarkWare Industries}. It has a large and growing community around it. The main purpose of Cairo is to be the language for creating provable programs. Cairo cannot verify an application since it is only the language for creating proofs. It can be extended with the Cairo-SHARP functionality to send the proof to an external server, which will then verify the proof.

\paragraph{Libiop.} This library provides zk-SNARK constructions that are transparent and post-quantum, which rely only on lightweight symmetric cryptography~\cite{Ben-Sasson&19}. It uses IOPs based on designs from the tools Ligero~\cite{Ames&17}, Aurora~\cite{Ben-Sasson&19}, and Fractal~\cite{Chiesa&20}. The tool creates zk-SNARKs with IOPs via a BCS transformation (Ben-Sasson-Chiesa-Spooner transformation~\cite{Ben-Sasson&16}). The BCS transformation preserves proof of knowledge and zero knowledge of the proof based on the behaviour of the IOP. The library is labelled as being \textit{not production ready} and having a somewhat slow issue-handling cycle. The latest activity is from April 2022, so it would suggest that still, some maintenance is happening in the library.

\paragraph{Bulletproofs.} The Bulletproofs library implements the protocol in~\cite{Bunz&18} and provides implementations of both single-party proofs of single or multiple ranges and online multi-party computation for range-proof aggregation. There is a programmable R1CS API for expressing, proving, and verifying proofs of arbitrary statements. The tool is not production ready, and the R1CS API feature is marked as experimental and unstable. The main benefit of Bulletproofs is the short proof size (logarithmic in witness size) without a trusted setup. Proof generation and verification times are linear.

The library provides the fastest implementation of the Bulletproof protocol that we could find and is written purely in Rust. The tool uses elliptic curves and relies on the discrete logarithm assumption making it vulnerable to quantum adversaries. The original paper's authors, however, propose a solution to this dilemma.

\paragraph{Zilch.} This framework is for deploying transparent zero-knowledge proofs without trusted third parties~\cite{Mouris&21}. The framework utilises zk-STARKs and relies on collision-resistant hash functions and the random oracle model, making it post-quantum resilient. The framework uses a custom abstract machine (zMIPS) adopted from the general-purpose MIPS processor architecture (Multi-party Interactive Proof System) to create arithmetic circuits and enable zero-knowledge proofs. Zilch uses a custom, high-level language called ZeroJava (a subset of Java) that supports object-oriented programming, albeit with a few limitations. A cross-compiler from ZeroJava to zMIPS and an assembly optimiser can be found in the repository.

\paragraph{LEGO-SNARK.}  This implementation is a library that works off the premise of having modular SNARK designs\cite{Campanelli&19}. It runs off a general composition tool that creates zk-SNARKS. It has the option via an additional layer of \textit{lifting tools} and \textit{gadgets} that it uses to be modular in the way that it can exchange the way QAPs are made and operate akin to that of other known zk-SNARK's.

\subsection{Experimental Evaluation of Tool Usability}

To assess the usage of the tools as off-the-shelf, we tried all of them by installing the latest version on stock hardware and ran a batch of tests/checks to assess several criteria.  We followed the steps prescribed for each installation in a Linux OS based on the Ubuntu distro (version 16.04 to 20.04). All tools were run from a docker container to provide a standard benchmark for easiness of installation and to keep the running environment controlled, increasing the possibility of reproducing the experiments. The stock hardware comprised an M1 chip laptop and a standard i7-based platform, which did not affect the experiences at this stage due to containerisation. 

\paragraph{Evaluation Criteria.} The criteria used to judge each tool is presented in the first column of Table \ref{tab:rank}. The first judgement was made on its \emph{Compilation of Framework} and to what extent a user can run the framework (from impossible to use due to unmet dependencies to a push button container execution). The \emph{Documentation}, \emph{Ease of Entry}, \emph{Community} and \emph{Support} criteria provide an indicator to what degree the user will find help in overcoming the typical issues encountered when adopting a tool (e.g., finding troubleshooting instructions for an error). Typically, adopting a tool leads to the need to adapt it to the particular problem it is being applied to. As a result, the user needs to have a good grasp of the tool's inner workings to obtain a practical solution to that problem. Thus, judging the degree of \emph{Complexity} of a tool allows a future user to estimate how easy it will be to adapt such tool examples to their particular computation scenario. The \emph{Maturity} criteria were used to judge how the protocols had grown since their original publication. It clarified which tools were being maintained and updated and which had not matured since the paper's initial publication. The \emph{Dependencies} criteria provide insight into the coupling the codebase had. With a lot of external dependencies comes even more maintenance. Finally, the \emph{Expressiveness} criterion judges the capabilities of the tools. Whether they support complex operations such as flow control and memory management or if it was simple compile-time operations only.


\begin{table}[ht!]
  \centering%
  \resizebox{\columnwidth}{!}{%
    \begin{tabular}{%
      >{\columncolor[HTML]{EFEFEF}}c ccccc}%
      \toprule%
      \textbf{Rank} &%
      \cellcolor[HTML]{EFEFEF}\textbf{1} &%
      \cellcolor[HTML]{EFEFEF}\textbf{2} &%
      \cellcolor[HTML]{EFEFEF}\textbf{3} &%
      \cellcolor[HTML]{EFEFEF}\textbf{4} &%
      \cellcolor[HTML]{EFEFEF}\textbf{5} \\ \midrule%
      \textbf{Compilation of Framework} &%
      It could not be compiled &%
      \begin{tabular}[c]{@{}c@{}}Many problems \\ when compiling.\\ A lot of manual\\ adjustments needed\end{tabular} &%
      \begin{tabular}[c]{@{}c@{}}Problems when compiling,\\ and few to non adjustments\\ to files needed\end{tabular} &%
      \begin{tabular}[c]{@{}c@{}}Minor problems,\\ and few to none \\ adjustments to \\ files needed\end{tabular} &%
      \begin{tabular}[c]{@{}c@{}}Easy compilation.\\ Potentially even a\\ Dockerfile\end{tabular} \\%
      \textbf{Documentation} &%
      No documentation &%
      \begin{tabular}[c]{@{}c@{}}Little and vague\\ documentation\end{tabular} &%
      \begin{tabular}[c]{@{}c@{}}Documentation with\\ sufficient descriptions\end{tabular} &%
      \begin{tabular}[c]{@{}c@{}}Wiki with a lot\\ of documentation\end{tabular} &%
      \begin{tabular}[c]{@{}c@{}}Wiki and related\\ guides. Continuously\\ updated.\end{tabular} \\%
      \textbf{Ease of Entry} &%
      \begin{tabular}[c]{@{}c@{}}No clear instructions\\ or examples on how\\ to create new proofs\end{tabular} &%
      \begin{tabular}[c]{@{}c@{}}Vague instructions\\ with no examples \\ on how to create new \\ proofs\end{tabular} &%
      \begin{tabular}[c]{@{}c@{}}Complex instructions with\\ no examples on how to\\ create new proofs\end{tabular} &%
      \begin{tabular}[c]{@{}c@{}}Somewhat simple\\ instructions with\\ examples on how to\\ create new proofs\end{tabular} &%
      \begin{tabular}[c]{@{}c@{}}Easy examples with\\ instructions on how\\ to create new proofs\end{tabular} \\%
      \textbf{Community} &%
      No community &%
      No active community &%
      \begin{tabular}[c]{@{}c@{}}Small community with\\ very limited activity\end{tabular} &%
      \begin{tabular}[c]{@{}c@{}}Some activity but it\\ might take time to \\ get help\end{tabular} &%
      \begin{tabular}[c]{@{}c@{}}Active community.\\ Possible to get help\\ and communicate \\ ideas\end{tabular} \\%
      \textbf{Support} &%
      \begin{tabular}[c]{@{}c@{}}No support and not\\ possible to contact\\ developers\end{tabular} &%
       &%
      \begin{tabular}[c]{@{}c@{}}Minimal support, an\\ associated email to contact\\ is given\end{tabular} &%
       &%
      \begin{tabular}[c]{@{}c@{}}Easy to get in contact\\ with. Active support\\ with forums or discord\end{tabular} \\%
      \textbf{Complexity} &%
      \begin{tabular}[c]{@{}c@{}}Code base is very hard\\ to understand\end{tabular} &%
      Hard to understand &%
      Somewhat easy to understand &%
      Easier to understand &%
      Easy to understand \\%
      \textbf{Maturity} &%
      \begin{tabular}[c]{@{}c@{}}No handling of issues\\ and pull requests\end{tabular} &%
      \begin{tabular}[c]{@{}c@{}}Acknowledges issues\\ in documentation\end{tabular} &%
      Must have issue handling &%
      \begin{tabular}[c]{@{}c@{}}Issues, but have \\ been acknowledged\end{tabular} &%
      \begin{tabular}[c]{@{}c@{}}Very few issues, and \\ issues are actively\\ being handled\end{tabular} \\%
      \textbf{Dependencies} &%
      \begin{tabular}[c]{@{}c@{}}Many required, external\\ dependencies.\end{tabular} &%
      \begin{tabular}[c]{@{}c@{}}Some required external\\ dependencies\end{tabular} &%
      \begin{tabular}[c]{@{}c@{}}Few required external\\ dependencies\end{tabular} &%
      \begin{tabular}[c]{@{}c@{}}Optional dependencies\\ or 1-3 required ones\end{tabular} &%
      No external dependencies \\%
      \textbf{Expressiveness} &%
      Special purpose &%
      Pure &%
      Stateful &%
      General control flow &%
      OOP \\ \bottomrule%
    \end{tabular}%
  }%
  \caption{Criteria used to rank the usability of the different tools in terms of its COTS readiness\label{tab:rank}}%
\end{table}

\subsection{Results}

Our hands-on experiments are in line with the accounts in the literature, where though full compiler toolchain implementations exist, they are somehow experimental and too brittle to be deemed general computation ready.\footnote{As of November 2022, a lot of the issues stem from the fundamental way the different libraries have been written: they are not modular and not easy to apply to other computational programs. Many libraries are based on LibSNARK, which provides good front and back ends (if both are chosen from libSNARK), but other implementations fall short because of this issue.}  The breakdown of the results of our experiments with each of the tools can be consulted in Table \ref{tab:results}, and in the following, we elaborate on the contents grouped by each line/tool.  Unless mentioned directly in the separate paragraphs, all of the tools were labelled as either not production ready or for experimental use.

\paragraph{Pequin.} The \textit{compilation} of this tool could be left to a shell script that would use an existing \textit{Dockerfile} to set up a container with the many dependencies required for the tool.  A well-documented setup guide helped with the navigation of the library and the creation of new verifiable programs. This made the more essential parts of \textit{Pequin} easier to understand immediately.  A moderate-sized community is available should one encounter issues or have questions regarding the tool. Should this not suffice, the administrating team behind the pepper project encourages one to contact them at their mail address for further help with their tool.  Many working examples in the library highlight that the tool supports many distinct high-complexity programs, even some requiring memory management.

\paragraph{Cairo.} The tool can be obtained through Python's package manager \textit{Pip}. To set up a working environment for \textit{Cairo}, a \textit{Dockerfile} was available in the GitHub repository, although we experienced some issues with it on non-UNIX systems. \textit{Cairo} is still being developed, and the team behind it just recently posted about releasing its first stable production-ready environment. This means that a developer might encounter bugs along the way. Thankfully, \textit{Cairo} has a large community that covers both a forum and a discord community with over 32.000 members. Besides these points, \textit{Cairo} provides excellent documentation on how the language can be used to develop programmable proofs, with many examples in their playground. Verifiable programs are created via the \textit{Cairo} compiler, which takes a \textit{.cairo} file with some main functionality. It is worth noting that proof cannot be verified locally on the client side. The proof must be sent to an external server known as SHARP to verify a computation.


\paragraph{Libiop.}

This tool is an implementation that produces transparent, post-quantum and symmetrically encrypted proofs, with the use of Interactive oracle proofs based on three internal tools: \textit{Ligero}, \textit{Aurora} and \textit{Fractal}.  Installing the library had its initial issues. \textit{Gtest}, which is Google's test framework, was from an older release that did not support the compilation of the release version of the library. The solution was to update \textit{Gtest} to a newer release.  Testing the individual tools was straightforward since several compiled tests for the different internal tools were available as bash scripts on the compilation. Though this did not give a clear insight into how the library was set up and if it even offered the option to create new proofs with the library.

\paragraph{Bulletproofs.} A pure Rust implementation by Dalek cryptography was tested. Initially, there were a few versioning issues with the tool's setup. Attempting to run the given test suite produced errors such as ``attributes are not yet allowed on if-expressions''. A \textit{fix} was to change the enforced nightly build in the \textit{rust-toolchains} file to a current version. This, however, had some side effects, as certain macros and methods used in the library were now flagged as deprecated. The deprecated functionality was manually updated to modern standards in the required files.
The tool is well-documented. The codebase is thoroughly commented on, and the developers have provided a wiki that describes structs, modules, etc., used in the implementation. Additionally, the mathematics behind the protocol is described in detail with a change in notation from the original Bulletproofs paper.  The library does not have a lot of dependencies. Using rustup to install rustc, cargo, and other standard Rust tools was sufficient for us. Creating and running new programs heavily resembles the traditional Rust process.

\paragraph{Zilch.} After installing the required dependencies, the tool was easy to start with as the Readme thoroughly describes how to use the ZeroJava compiler, the compilation process, and the zMIPS extension to the MIPS ISA.  Unfortunately, the documentation related to the syntax of ZeroJava is not given the same level of attention to detail. Small things like not being able to add or multiply more than two integers without including parentheses, or being unable to add two array values together in a single line, are not mentioned in the documentation. Also, sometimes not defining new variables at the top of the program caused compilation to fail.  Despite these inconsistencies, Zilch is one of the few tools that support object-oriented programming. However, ZeroJava does have a few limitations compared to modern object-oriented languages. For instance, it does not support interfaces or function overloading; it only has a single inheritance. Furthermore, constructs such as for-loops and 2D arrays are not supported.  It is also worth noting that some of the tests in the repository might not work correctly. For instance, the algorithm used for matrix multiplication is flawed and failed to verify. The matrix multiplication failed to verify even after adjusting the algorithm so that the correct matrix product was computed.

\paragraph{LEGO-SNARK.}
The main problem with this tool was that the dependency on libSnark was from an older release that was no longer compatible with the implementation of the current state of the tool. Therefore it was required to manually go in and change the release version of libSnark to the newest release so that it would work with the tool.  Running examples in the library was straightforward after the Makefile had run its magic, though few examples were present.  The code itself did not give a very clear inside of what was happening in each example. Therefore recreating proofs or creating new ones is rather complicated.  There was also an issue with a utility file. Due to a deprecated function from an external library, the \textbf{util} code had to be updated to be able to compile the examples.  From the paper, LEGO-SNARK sounds smart since it has a modular SNARK design, but in practice, this is not intuitive to work with. We had a hard time understanding how to extend the library with new functions since the repository or the paper does not go into depth about how the original author got to the example proofs that the library holds.

\subsection{Discussion}

Overall, none of the works provides an off-the-shelf framework. For the surveyed tools, an adopter has to build its client-server architecture and reuse components to implement the proof system or derive the program specification $\clz \Sigma$. Even in cases where the toolchain provides a $\clz \Sigma$, we experienced that it is easy to find unsupported corner cases where the compiler cannot find a $\clz \Sigma$ for a simple C routine making additions, for example. Below we break down our assumptions and conclusions after using each tool.

\paragraph{Pequin.} The idea of Pequin was that it was supposed to be an easier implementation of the original library titled Pepper. A docker file was available for easier implementation, together with several bash scripts to fetch dependencies and READMEs to help with both the installation and the actual testing of the tool. Pequin delivered an easy setup with a very user-friendly way to start making your proofs. \\

\paragraph{Cairo.} The idea of Cairo was to create a new language from scratch, which would be utilised to make provable programs. It also had a connected verification server in the form of Cairo-SHARP, which would allow for verification of the provable programs.

\paragraph{Libiop.} The idea of libiop was that it would utilise interactive oracular proofs. This would help make the libraries quantum-safe, transparent and lightweight. It consisted of 3 internal libraries (Ligero, Aurora and Fractal) with a test program. Working with libiop was not very informative as to how tests were made; it was obvious to see the tests run and the computation being proved and verified. When trying to create new proofs, it becomes much more complicated. It is not an intuitive setup to create new proofs and can therefore require a bit of further reading.

\paragraph{Bulletproofs.} The primary goal of Bulletproofs was to allow for efficient range proofs on committed values -- a type of problem that is not limited to but occurs frequently in blockchain-based cryptocurrencies. The proof protocol has short proofs that are only logarithmic in the witness size and supports multiparty computation. Beyond range proofs, the tool also computes inner product proof.  The underlying protocol is only designed to do a limited number of things, so unlike other tested tools, Bulletproofs only allow for high-level implementations of range and inner product proofs. To prove and verify arbitrary statements, one must use the R1CS API, which is flagged as experimental. However, creating and running programs requires little domain-specific knowledge, especially if one is familiar with Rust.

\paragraph{Zilch.} The main idea of the Zilch tool is to abstract the encoding of computations as arithmetic circuits away from the user as debugging and optimising programs, as circuits are unfamiliar to most users and require domain-specific knowledge. Instead, Zilch is one of the few tools that uses a high-level language that supports an object-oriented programming style.  The ZeroJava compiler in the repository creates and runs new programs similar to compiling and running a regular Java program. However, syntactically the parser was very strict and not all the behaviour is documented. Furthermore, not all the tests in the repository worked as intended -- like the matrix multiplication. Specific programs are a hassle to write in ZeroJava due to the number of inconsistencies hinted at in the result section and the lack of constructs such as for-loops and 2D arrays.

\paragraph{LEGO-Snark} The idea of LEGO-Snark is to have a modular design, which allows for interchangeable SNARK protocols.  After compiling the library, a few examples were available for testing. Looking through the code of the examples made it clear that only three commands were used to produce the proof and the verification. Still, a guide on using these functions was not present.


\begin{table}[ht]
  \resizebox{\columnwidth}{!}{%
    \begin{tabular}{@{}llrrrrrrrrrr@{}}
      \toprule
      \rowcolor[HTML]{EFEFEF}
      \multicolumn{1}{l}{\textbf{\begin{tabular}[c]{@{}l@{}}Developer\end{tabular}}} &
      \multicolumn{1}{l}{\textbf{\begin{tabular}[c]{@{}l@{}}Tool\end{tabular}}} &
      \multicolumn{1}{l}{\textbf{\begin{tabular}[c]{@{}l@{}}Comp-\\ ilation\end{tabular}}} &
      \multicolumn{1}{l}{\textbf{\begin{tabular}[c]{@{}l@{}}Documen-\\ tation\end{tabular}}} &
      \multicolumn{1}{l}{\textbf{Entry}} &
      \multicolumn{1}{l}{\textbf{\begin{tabular}[c]{@{}l@{}}Com-\\ munity\end{tabular}}} &
      \multicolumn{1}{l}{\textbf{\begin{tabular}[c]{@{}l@{}}Sup-\\ port\end{tabular}}} &
      \multicolumn{1}{l}{\textbf{\begin{tabular}[c]{@{}l@{}}Com-\\ plexity\end{tabular}}} &
      \multicolumn{1}{l}{\textbf{\begin{tabular}[c]{@{}l@{}}Matu-\\ rity\end{tabular}}} &
      \multicolumn{1}{l}{\textbf{\begin{tabular}[c]{@{}l@{}}Depen-\\ dencies\end{tabular}}} &
      \multicolumn{1}{l}{\textbf{\begin{tabular}[c]{@{}l@{}}Expres-\\ siveness\end{tabular}}} &
      \multicolumn{1}{l}{\textbf{Overall}} \\ \midrule
      \multirow{ 3}{*}{Pepper Project}
      & Ginger & 3 & 2 & 3 & 3 & 3 & 3 & 3 & 2 & 4 & \textbf{2.89} \\
      & Zaatar & 3 & 2 & 3 & 3 & 3 & 3 & 3 & 2 & 4 & \textbf{2.89} \\
      & Pequin & 5 & 4 & 5 & 4 & 3 & 5 & 4 & 2 & 4 & \textbf{4.00} \\ 
      StarkWare Industries & Cairo & 5 & {\color[HTML]{330001} 5} &5 &5 &5 &4 &3 &3 &4 &\textbf{4.33} \\
      Scipr-lab &Libiop &4 &3 &3 &4 &3 &3 &3 &3 &4 &\textbf{3.33} \\
      Dalek cryptography & Bulletproofs & 3 & 4 & 4 & 4 & 3 & 4 & 4 & 4 & 4 &\textbf{3.78} \\ 
      Trustworthy Computing Group & Zilch & 4 & 2 & 4 & 1 & 1 & 2 & 1 & 2 & 4.5 & \textbf{2.39} \\ 
      IMDEA-software & LEGO-SNARK & 3& 3 & 2 & 1 & 3 & 2 & 3 & 3 & 3 & \textbf{2.56}  \\\bottomrule
    \end{tabular}%
  }%
  \caption{Implementation scores for the evaluation criteria.\label{tab:results}}
\end{table}

\section{Conclusions}\label{sect:conclusions}

It is not easy to check a mathematical proof~\cite{Sudan&09}. We may suspect an error but have no counterexample to use to reject the proof. Scanning through the proof, flipping the pages randomly and reading a line here and there would be ideal. But could we ever be confident that we could find the errors in such a superficial way? The usual presentation of proof is not like this. It is easy to write proof of a false assertion where the mistake is subtly hidden.

But we do not have to present proofs traditionally. The last 25 years have seen the development of new formats for writing mathematical proofs. These new formats rely on profound results in theoretical computer science. These results underpin probabilistic algorithms for validating the proofs of general assertions. A checker does not need to review an entire proof. Instead, they need to read only a constant number of bits to gain confidence in the proof. Someone who proves a theorem can rewrite the proof in the new format. They can then give it to the checker with the assurance that the reviewer will believe this new proof. A malicious person might claim a false assertion is a theorem. Since it is false, there can be no proof of its correctness. If they propose a pretend proof, the checker will discover an error with near certainty. It does not matter whether the so-called proof is in the new format.

This report has surveyed the state of the art and summarised the theoretical foundations that underlie this result. This includes the PCP theorem. It also provides for the notion of zero-knowledge proofs used to shield clients working with PCP, with a summary of recent constructions for deploying zero-knowledge proof systems in practice. Remarkably, a statement can be proven correct in zero-knowledge with such short proofs and efficient prover and verifier algorithms. As Sudan points out~\cite{Sudan&09}, PCPs make verification more robust.

Our work also surveyed the practice in the field of verifiable computations. We tested several implementations of the state-of-the-art proof systems put forward in recent research papers and ranked them according to several criteria to evaluate whether a development team can use the implementation as a component off-the-shelf, i.e., without developing the implementation of the proof system itself.  Overall, none of the implementations found provides an off-the-shelf framework. When using an implementation, the user has to build its client-server architecture. The most advanced implementations provide a toolchain compiling user-defined programs into a prover and verifier. Still, we experienced that finding unsupported corner cases where the toolchain cannot deal with a specific program is easy.

Our test results indicate that Pequin, a tool that refactored several tools~\cite{WalfishB2015} and packaged them into a single toolchain, is an excellent candidate to transform into a verifiable computation framework. In particular, we asserted that.  Pequin is currently being developed and packaged in a container ready to use. Nevertheless, there are still some pieces to be put together if the codebase is to be a ready-to-use component. Another tool faring well in our experiments is Cairo. It is also being developed, and we consider it another good candidate for choice as an off-the-shelf component. Our survey covered some tools that are not being actively developed: Libiop, Zilch and
 Bulletproofs, but the research results they illustrate are interesting from a user perspective.  For instance, Libiop reports highly efficient and transparent proofs that may interest our application domain. LEGO-SNARK was chosen because its modularity allows different implementations of the zk-SNARKS in the back end. Nevertheless, only Libiop and Bulletproofs fare well in our ranking.

\newpage%

\bibliographystyle{plain} 
\bibliography{simon,diego,jim-bib,stefan,tools}

\end{document}